\documentstyle[12pt,emlines2]{article}
\begin{document}
\title{
\begin{flushright}
{\small SMI-25-96 }
\end{flushright}
\vspace{2cm}
Incidence Matrix Description of \\ Intersecting  p-brane Solutions }
{\it\author{
I.Ya.Aref'eva
\thanks
{ e-mail: arefeva@arevol.mian.su}
\\
Steklov Mathematical Institute,\\
Vavilov 42, GSP-1, 117966, Moscow, Russia\\
and\\
O.A.Rytchkov
\thanks
{ e-mail: rytchkov@grg1.phys.msu.su}\\
Department of Theoretical Physics,\\
Moscow State University,\\
Moscow 119 899, Russia}}
\date {$~$}
\maketitle
\begin {abstract}
An algebraic method for a general construction of intersecting p-brane
solutions in diverse spacetime dimensions is discussed. An incidence
matrix describing configurations of electric and magnetic fields is
introduced. Intersecting p-branes are specified by solutions of a system
of characteristic algebraic equations for the incidence matrix. This
set of characteristic equations generalizes a single characteristic
equation found before for a special "flower" ansatz. The  characteristic
equations admit solutions only for quantized values of  scalar coupling
parameters. A wide list of examples including solutions with regular
horizons and non-zero entropy in D=11 and D=10 theories is presented.
\end {abstract}

\newpage
\section{Introduction}
\setcounter{equation}{0}

Soliton solutions in $D=10$ superstring theories and $D=11$ supergravity
support an idea that five known $D=10$ superstrings are in fact five
different points of an unique $M$-theory \cite{HT}-\cite{JHS}.
In this context
it is interesting to understand whether an existence of soliton
solutions is a privilege  of effective superstring theories  or
$D=11$ supergravity. For this purpose it is useful to have a list
of soliton solutions in $D$-dimensional theory with the following
action
\begin{equation}
 I=\frac{1}{2\kappa ^{2}}\int d^{D}x\sqrt{-g}
 (R- \frac{1}{2}(\nabla \phi)^{2}-\frac{e^{-\alpha \phi}}
{2(d+1)!}F^{2}_{d+1}),
                                           \label{1}
 \end{equation}
where $F_{d+1}$ is a $d+1$ differential form,
 $F_{d+1}=d{\cal A}_{d}$, $\phi$ is a dilaton.
In general, the fields $F$ and $\phi$ are  linear combinations of a
large variety of gauge fields and scalars.
\par

The aim of this paper is to present a systematic description
of a class
of soliton solutions in higher dimensional gravity coupled with matter.
This class describes multi-soliton composite p-brane solutions.
Elementary p-brane solutions
have been found years ego
\cite{DGHR}-\cite{Re} (see also \cite{Lu}-\cite{Cvetic}).
The recent microscopic interpretation
of the Bekenstein-hawking entropy within string theory \cite{Vafa}-\cite{CM}
has stimulated an investigation of p-brane solutions \cite{TS}-\cite{IN}.
Composite p-brane solutions for $D=11$ and $D=10$  have been
obtained recently ~\cite{TS,Ts1,ToPa1}.
We will also search
for  an universal higher dimensional (higher then D=11) theory  which
has p-brane solutions that after dimension reduction could  produce all
known $D=11$ and $D=10$ p-brane solutions.
\par

To get a wide class of composite p-brane solutions
with non-zero electric and magnetic
charges
in theory (\ref{1})
we will use an algebraic method \cite{AV,AVV,IN},
which permits to reduce the problem of finding solutions to an algebraic
problem.
\par
We will describe the class of solutions with a $N+2$-blocks metric and
a gauge field ${\cal A}$ which has
$E$ electric and $M$ magnetic components (branches)

\begin{equation}
                                         \label{3'}
{\cal A}=\sum _{a}^{E}{\cal A}^{(\cal E)}_{a}+
\sum _{b}^{M}{\cal A}^{(\cal M)}_{b}.
\end{equation}

An $N+2$-blocks metric has the following form
\begin{equation}
ds^{2}=e^{2A(x)}\eta _{\mu \nu} dy^{\mu}
dy^{\nu}+\sum _{i=1}^{N}e^{2F_{i}(x)}dz_{i}^{m_{i}}dz_{i}^{m_{i}}
+e^{2B(x)}dx^{\gamma}dx^{\gamma},
                                          \label{3}
\end{equation}
where $\eta_{\mu\nu}$
is a flat Minkowski metric, $\mu$, $\nu$ run from $0$ to $q-1$,  $m_{i}$
runs
 from $1$ to $r_{i}$,
$\gamma$ runs from
$1$ to $s+2$ and
\begin{equation}
                                         \label{3''}
D=q+\sum _{i}^{N}r_{i}+s+2,
\end{equation}
In (\ref{3}) $A$, $B$ and $F_{i}$ are functions of $x$.
\par
To describe  electric (magnetic)
configurations ${\cal A}^{(\cal E)}_{a}$
(${\cal A}^{(\cal M)}_{b}$)  we will introduce
 an electric (magnetic) {\it incidence} matrix. The incidence matrix, say the
electric
incidence matrix, is an rectangular $N\times E$ matrix,
\begin{equation}
                                      \label{4}
\Delta =(\Delta _{ai}), ~~a=1,...E, ~~i=1,...N,
\end{equation}
which  rows
correspond to independent branches of  electric gauge field
 and  columns refer to the number of the blocks in the
 metric (\ref{3}). The entries
of the incidence matrix are equal to 0 or 1.
 Field configurations  ${\cal A}^{(\cal E)}_{a}$  will be in the
following form

\begin{equation}
                                      \label{5}
{\cal A}^{(\cal E)}_{a}=h_a H_{a}^{-1}\omega_0
\prod _{\{i|\Delta _{ai}=1\}}\wedge \omega_i,
\end{equation}
where $\omega_0$ and $\omega_i$ are  $q$ and  $r_i$-form, respectively,

$$\omega_0=dy^0\wedge\ldots
  \wedge dy^{q-1}, ~~~\omega_i=dz^1_i\wedge\ldots\wedge dz^{r_i}_i.$$
$H_{a}$ are some functions of $x$-variables, $a=1,\ldots,E$.

 We will show that in the case of a pure electric ansatz,
\begin{equation}
                                         \label{6'}
{\cal A}=\sum _{a}^{E}{\cal A}^{(\cal E)}_{a},
\end{equation}
the following metric
$$ds^{2}=(H_{1}H_{2}...H_{E})^{-2t\sigma }
\sum _{\mu , \nu =0}^{q-1}\eta_{\mu \nu} dy^{\mu} dy^{\nu}+$$
\begin{equation}
(H_{1}H_{2}...H_{E})^{2u\sigma }[
\sum _{i}(\prod _{a}H_{a}^{-\Delta _{ai}})^{\sigma}
\sum _{m_{i}=1}^{r_{i}}dz_{i}^{m_{i}}dz^{m_{i}}
+\sum _{\gamma }dx^{\gamma}dx^{\gamma}],
                                              \label{6}
\end{equation}
is a solution of the theory (\ref{1}).
Here
\begin{equation}
                                       \label{6''}
 u=\frac{d}{2(D-2)},~~t=\frac{1}{2}-u,~~
\sigma =\frac{1}{dt+\alpha
^{2}/4},
\end{equation}
$H_{1},H_{2}...,H_{E}$ are arbitrary harmonic functions  of
x-variables  and
 the
incindence matrix $\Delta $  satisfies the following $E(E-1)/2$
{\it characteristic} equations

\begin{equation}
                                       \label{7}
(q+\frac{\alpha
^{2}}{2})-\frac{d^2}{D-2}+\sum_{i=1}^{N}r_i\Delta_{ai}\Delta_{a'i}=0,
~~a\neq a'.
\end{equation}
Note that the characteristic equations (\ref{7}) generalize a characteristic
algebraic equation for a so-called "flower' ansatz \cite{IN}.
In this ansatz   $\sum_{i=1}^{N}r_i\Delta_{ai}\Delta_{a'i}=0,$ and one has
a single equation  \cite{IN}

\begin{equation}
                                       \label{7a}
(q+\frac{\alpha
^{2}}{2})=\frac{d^2}{D-2},
\end{equation}

In the case when  electric and
magnetic components of the d-form gauge field are non-zero (see eqs.
(\ref{1.2}) below) the metric solving the theory (\ref{1}) has the
form

$$ds^{2}=
(H^{-1}_{1}H^{-1}_{2}...H^{-1}_{E}U_{1}U_{2}...U_{M})^{2t\sigma}
(H_{1}H_{2}...H_{E})^{\sigma}\times $$

$$
\{ (H_{1}H_{2}...H_{E}U_{1}U_{2}...U_{M})^{-\sigma}
\sum _{\mu, \nu =0}^{q-1} \eta_{\mu \nu}dy^{\mu} dy^{\nu}+$$
\begin{equation}
 (U_{1}U_{2}...U_{M})^{-\sigma}\sum _{i}(\prod
_{a}H_{a}^{-\Delta _{ai}} \prod _{b}U_{b}^{\Lambda _{bi}})^{\sigma}\sum
_{m_{i}=1}^{r_{i}}dz_{i}^{m_{i}}dz_{i}^{m_{i}}
+\sum _{\alpha
}dx^{\alpha}dx^{\alpha}\},
                                    \label{8}
\end{equation}
where $H_{a}$ and $U_{b}$ are arbitrary harmonic
 functions of x-variables, $a=1,..,E$, $b=1,..,M$, and  the electric
and magnetic incidence matrix, $\Delta _{ai}$ and  $\Lambda _{bi}$,
satisfy two more conditions
\begin{equation}
                                    \label{9}
s+{\alpha^2\over
2}-\frac{d^2}{D-2}+\sum_{i=1}^Nr_i\Lambda_{bi}\Lambda_{b'i}=0,
b\neq b',~~b,b'=1,..M
\end{equation}

\begin{equation}
                                       \label{10}
{\alpha^2\over
2}-\frac{d^2}{D-2}+\sum_{i=1}^Nr_i\Delta_{ai}\Lambda_{bi}=0,
~~a=1,...E,~~b=1,..M
\end{equation}

The  characteristic
equations admit solutions only for quantized values of  scalar coupling
parameters  that is in accordance with \cite{AV,Berg,IN}.
Note that the case when $\alpha =0$ (D=11 supergravity)
p-brane solutions are given by the  above formula with $\alpha =0$.
These equations are very restrictive since they must be solved for integers
and they admit non-trivial solutions only for special $D$.
The selected $D$ are the same as for the  "flower"  ansatz \cite{IN}.
In particular, they are $D=10,~ 11,$ $18,~ 20,~ 26$.

For the case of two ($k$) gauge fields we will  introduce four ($2k$)
incidence
matrices and the
corresponding expressions for metric
will be  very similar (see eqs.(\ref{genmetr})
below). The corresponding incidence matrix will satisfy to relations
similar to
(\ref{7}), (\ref{9}) and (\ref{10}).

The algebraic method \cite{AV,AVV,IN} can be also used for
finding solutions with depending harmonic functions.
In particular,
one can consider a generalized dyonic ansatz
that contain equal numbers of magnetic and electric branches,
${\cal E}={\cal M}$ and the same set of harmonic functions
$H_{i}=U_{i}$, $i=1,...{\cal E}$. This construction gives in
$D=11 $  a new dyonic solution with regular horizon and non-zero entropy.

\par
The paper is organized as follows. In Section 2 we present soliton
solutions for the theory (\ref{1}) with one d-form field with dilaton
and in  Section 3 we present soliton
solutions for a theory  with two antisymmetric  fields
and dilaton. Both sections start from general considerations and
are concluded by some examples. There one can also find relations with known
solutions. In section 4 we collect some examples of solutions of the form
(\ref{8}) with depending functions
$H_{1}, H_{2}...H_{E}$ and $U_{1}U_{2}...U_{M}$.

\section{Gravity Coupled with $d$-form and Dilaton }
\subsection{General Consideration}
\setcounter{equation}{0}
Let $M^D$ be a manifold with a following structure:
$$M^D={\bf M}^q\times{\bf R}^{r_1}\times\cdots\times{\bf R}^{r_N}
\times{\bf R}^{s+2},$$
and the metric has the form (\ref{3}).

For the d-form we are going to consider a class of ansatzes corresponding
to E electric and M magnetic charges. Each ansatz we
will connect with two rectangular matrices
$\Delta_{ai},\;a=1..E,\;i=1..N$ and $\Lambda_{bi},\;b=1..M,\;i=1..N$
with the following properties:
\par
1) $\Delta_{ai}=1$ or $0$;
\par
1') $\Lambda_{bi}=1$ or $0$;
\par

2) there are no $a_1$ and
  $a_2\;(a_1\not=a_2)$ such that: $\Delta_{a_1i}=\Delta_{a_2i}$ for
all $i$
\par
2') there are no $b_1$ and
  $b_2\;(b_1\not=b_2)$ such that:
$\Lambda_{b_1i}=\Lambda_{b_2i}$ for all $i$;
\par
3) there are no $a_1$ and
  $a_2\;(a_1\not=a_2)$
 and $i_1,\;i_2$ such that: $\Delta_{a_1i}=\Delta_{a_2i}$ for all
$i\neq i_1,\;i_2$ and $r_{i_1}=r_{i_2}=1$
\par
3') there are no $b_1$ and
  $b_2\;(b_1\not=b_2)$
 and $i_1,\;i_2$ such that: $\Delta_{b_1i}=\Delta_{b_2i}$ for all
$i\neq i_1,\;i_2$ and $r_{i_1}=r_{i_2}=1$
 \par

4)$\;\sum_{i=1}^N r_i\Delta_{ai}=d-q\,$ for all $a$;
\par
4')$\;\sum_{i=1}^N r_i\Lambda_{bi}=d-s\,$ for all $b$;
 \par
5)$\Delta_{ai}\not=\Lambda_{bi} $ if $s=1,\;q=1$
  \par
An ''electric'' field is assumed in the form

\begin{equation}
                                    \label{1.1}
{\cal  A}=\sum_{a=1}^{\cal E}h_ae^{C_a(x)}\omega _{0}^{q}
\prod _{\{i|\Delta _{ai}=1\}}\wedge~\omega_i
\end{equation}
where
$\omega _{0}^{q}$ ia a q-form
$$\omega _{0}^{q}=dy^0\wedge\ldots
  \wedge dy^{q-1}$$
 $\omega_i$ is a $r_i$-form
$$\omega_i=dz^1_i\wedge\ldots\wedge dz^{r_i}_i.$$

A "magnetic" field is assumed in the form
\begin{equation}
                                    \label{1.2}
F=\sum_{b=1}^Mv_be^{\alpha\phi+\chi_b(x)}\ast_bd\chi_b(x),
\end{equation}
where $\ast_b$ is a Hodge dual on
$${\bf M}_{b}^{d+2}=(\prod _{\{i|\Lambda _{bi}=1\}}{\bf
R}^{r_i}\times){\bf R}^{s+2}.
$$

We will refer to the different terms in (\ref{1.1}) and (\ref{1.2})
as to the different branches of electric and magnetic fields.
In order to simplify the Ricci tensor
\cite{AVV,AV,IN} we assume
\begin{equation} \label{1.3} qA+\sum
r_iF_i+sB=0. \end{equation} We get \begin{equation} \label{1.4}
R_{\mu\nu}=-\eta_{\mu\nu}e^{2(A-B)}\Delta A,
\end{equation}
\begin{equation}
                                    \label{1.5}
R_{m_in_i}=-\delta_{m_in_i}e^{2(F_i-B)}\Delta F_i,
\end{equation}
\begin{equation}
                                    \label{1.6}
R_{\alpha\beta}=-q\partial_\alpha A\partial_\beta A-
\sum_i r_i\partial_\alpha F_i\partial_\beta F_i-
s\partial_\alpha B\partial_\beta B-\delta_{\alpha\beta}\Delta B.
\end{equation}

 Components of the energy-momentum tensor for the  d-form gauge field,
 \begin{equation}
                                    \label{1.7}
T_{MN}^{(F)}={1\over 2d!}(F_{MM_1\ldots M_d}F_N^{M_1\ldots
 M_d}-{1\over 2(d+1)}g_{MN}F^2)e^{-\alpha\phi}
\end{equation}
 for the above ansatz are
 $$T_{\mu\nu}^{(F)}=-\eta_{\mu\nu}e^{2(A-B)}[\sum_{a=1}^E{h_a^2\over
  4}e^{-\alpha\phi-2qA-2\sum_{i=1}^N\Delta_{ai}r_iF_i+2C_a}(\partial
C_a)^2+$$
\begin{equation}
                                    \label{1.8}
 +\sum_{b=1}^M{v_b^2\over
  4}e^{\alpha\phi+2sB+2\sum_{i=1}^N\Lambda_{bi}r_iF_i+2\chi_b}(\partial
\chi_b)^2],
\end{equation}
 $$
T_{m_in_i}^{(F)}=\delta_{m_in_i}e^{2(F_i-B)}\lbrack\sum_{a=1}^E{h_a^2\over 4}
e^{-\alpha\phi-2qA-2\sum_{i=1}^N\Delta_{ai}r_iF_i+2C_a}(\partial
   C_a)^2(1-2\Delta_{ai})-$$
 \begin{equation}
                                    \label{1.9}
-\sum_{b=1}^M{v_b^2\over 4}
e^{\alpha\phi+2sB+2\sum_{i=1}^N\Lambda_{bi}r_iF_i+2\chi_b}(\partial
   \chi_b)^2(1-2\Lambda_{bi})\rbrack,
\end{equation}
$$T_{\alpha\beta}^{(F)}=-\sum_{a=1}^E{h_a^2\over
2}e^{-\alpha\phi-2qA-2\sum_{i=1}^N\Delta_{ai}
r_iF_i+2C_a}\lbrack\partial_\alpha C_a\partial_\beta
    C_a-{\delta_{\alpha\beta}\over 2}(\partial C_a)^2\rbrack-$$
  \begin{equation}
                                    \label{1.10}
\sum_{b=1}^M{v_b^2\over 2}e^{\alpha\phi+2sB+2\sum_{i=1}^N\Lambda_{ai}
r_iF_i+2\chi_b}\lbrack\partial_\alpha \chi_b\partial_\beta
    \chi_b-{\delta_{\alpha\beta}\over 2}(\partial \chi_b)^2\rbrack.
\end{equation}
 All others components are zero. Note that we assume 3) and 3') to guarantee
the diagonal form of the energy-momentum tensor.

Components of the energy-momentum tensor for dilaton
$$T_{MN}^{(\phi)}={1\over 2}(\partial_M\phi\partial_N\phi-{1\over
2}g_{MN}(\nabla\phi)^2)$$
 have the following non-zero components
\begin{equation}
                                    \label{1.11}
T_{\mu\nu}^{(\phi)}=-{1\over 4}\eta_{\mu\nu}e^{2(A-B)}(\partial\phi)^2,
\end{equation}
  \begin{equation}
                                    \label{1.12}
T_{m_in_i}^{(\phi)}=-{1\over
 4}\delta_{m_in_i}e^{2(F_i-B)}(\partial\phi)^2,
\end{equation}
  \begin{equation}
                                    \label{1.13}
T_{\alpha\beta}^{(\phi)}={1\over
 2}[\partial_\alpha\phi\partial_\beta\phi-{1\over
 2}\delta_{\alpha\beta}(\partial\phi)^2].
 \end{equation}
If we assume the following "no-force" conditions
 \begin{equation}
                                    \label{1.14}
-\alpha\phi-2qA-2\sum_{i=1}^N\Delta_{ai}r_iF_i+2C_a=0,\;a=1..E
\end{equation}

 \begin{equation}
                                    \label{1.15}
\alpha\phi+2sB+2\sum_{i=1}^N\Lambda_{bi}r_iF_i+2\chi_b=0,\;b=1..M
\end{equation}
the form of $T_{MN}$ crutially simplifies and the Einstein equations cast into
the  form

\begin{equation}
                                    \label{1.16}
\Delta A=\sum_{a=1}^Eth_a^2(\partial
C_a)^2+\sum_{b=1}^Muv_b^2(\partial\chi_b)^2,
\end{equation}
  \begin{equation}
                                    \label{1.17}
\Delta F_{i}=\sum_{a=1}^Eh_a^2
(\partial C_a)^2(u-{1\over 2}\Delta_{ai})-\sum_{b=1}^Mv_b^2
(\partial \chi_b)^2(u-{1\over 2}\Lambda_{bi}),
\end{equation}
$$
-q\partial_\alpha A\partial_\beta A-
\sum_i r_i\partial_\alpha F_i\partial_\beta F_i-
s\partial_\alpha B\partial_\beta B-\delta_{\alpha\beta}\Delta B$$
 $$
={1\over
2}\partial_\alpha\phi\partial_\beta\phi-\sum_{a=1}^Eh_a^2\lbrack{1\over
    2}\partial_\alpha C_a\partial_\beta C_a-u\delta_{\alpha\beta}
    (\partial C_a)^2-$$
\begin{equation}
                                    \label{1.18}
 \sum_{b=1}^Mv_b^2\lbrack{1\over
    2}\partial_\alpha \chi_b\partial_\beta\chi_b-t\delta_{\alpha\beta}
    (\partial \chi_b)^2\rbrack,
\end{equation}
where $t$ and $u$ are given by (\ref{6''}).

 Equation of motion for dilaton reads
\begin{equation}
                                    \label{1.19}
\triangle\phi-{\alpha\over 2}\bigl[\sum_{a=1}^Eh_a^2(\partial
C_a)^2-\sum_{b=1}^Mv_b^2(\partial \chi_b)^2\bigr]=0
\end{equation}

 From Maxwell's equations under conditions (\ref{1.14}) and
(\ref{1.15}) we get \begin{equation} \label{1.20} \Delta
C_a=(\partial C_a)^2 \end{equation}

 From Bianchi identity for $F$ under conditions (\ref{1.14}) and
(\ref{1.15})  we have
\begin{equation}
                                    \label{1.21}
\Delta \chi_b=(\partial\chi_b )^2,
\end{equation}
therefore, $H_{a}=e^{-C_{a}}$, $a=1,..,E$ and $U_{b}=e^{-\chi _{b}}$,
$b=1,...M$
 are harmonic functions.

We solve  (\ref{1.16}), (\ref{1.17}), (\ref{1.19}) and the diagonal part of
(\ref{1.18})
 supposing
 \begin{equation}
                                    \label{1.22}
A=\sum_{a=1}^Eth_a^2C_a+\sum_{b=1}^Muv_b^2\chi_b
\end{equation}
\begin{equation}
                                    \label{1.23}
 F_i=\sum_{a=1}^Eh_a^2C_a({1\over 2}\Delta_{ai}-u)-
\sum_{b=1}^Mv_b^2\chi_b({1\over 2}\Lambda_{bi}-u)
\end{equation}
\begin{equation}
                                    \label{1.24}
B=-u\sum_{a=1}^Eh_a^2C_a-t\sum_{b=1}^Mv_b^2\chi_b
\end{equation}
\begin{equation}
                                    \label{1.25}
\phi={\alpha\over 2}\bigl[\sum_{a=1}^Eh_a^2 C_a-
\sum_{b=1}^Mv_b^2 \chi_b\bigr]
\end{equation}

But now we have to guarantee relations (\ref{1.3}),
(\ref{1.14}), (\ref{1.15}) and non-diagonal part of
(\ref{1.18}). Let us analyze these conditions step by step.

 A substitution of (\ref{1.22})--(\ref{1.25}) in (\ref{1.14})
under an assumption of the  independence of $C_a$ and $\chi_b$ gives
three types of relations
\begin{equation}
                                    \label{1.26}
1={\alpha^2\over
4}h_a^2+qth_a^2+\sum_{i=1}^Nr_i\Delta_{ai}h_a^2({1\over2}\Delta_{ai}-u)
\end{equation}
\begin{equation}
                                    \label{1.27}
0={\alpha^2\over
4}+qt+\sum_{i=1}^Nr_i\Delta_{ai}h_{a}^2({1\over2}\Delta_{a'i}-u),
\;a\not=a'
\end{equation}
\begin{equation}
                                    \label{1.28}
0=-{\alpha^2\over
4}+qu-\sum_{i=1}^Nr_j\Delta_{ai}({1\over2}\Lambda_{bi}-u)
\end{equation}

Taking into account requirement 4) on the electric incidence matrix
from equation
(\ref{1.26}) we get
 \begin{equation}
                                    \label{1.29}
h_a^2=h^2=\sigma
\end{equation}
 where $\sigma $ is given by equation (\ref{6''}).

Equations  (\ref{1.27}) and (\ref{1.28})
together with conditions 4) and 4') give

 \begin{equation}
                                    \label{1.30}
(D-2)({\alpha^2\over
2}+q)-d^2+(D-2)\sum_{i=1}^Nr_i\Delta_{ai}\Delta_{a'i}=0,\;a\neq a'
\end{equation}

\begin{equation}
                                    \label{1.31}
(D-2){\alpha^2\over
2}-d^2+(D-2)\sum_{i=1}^Nr_i\Delta_{ai}\Lambda_{bi}=0,
\end{equation}
that are nothing but conditions (\ref {7}) and (\ref{10})
mentioned in the Introduction.

 A substitution of (\ref{1.22})--(\ref{1.25}) in (\ref{1.15})
under an assumption of the  independence of $C_a$ and $\chi_b$ gives
once again three types of relations. They are
\begin{equation}
                                    \label{1.32}
 1={\alpha^2\over
4}v_b^2+stv_b^2+\sum_{i=1}^Nr_i\Lambda_{bi}v_b^2({1\over2}\Lambda_{bi}-u)
\end{equation}

\begin{equation}
                                    \label{1.33}
0={\alpha^2\over
4}+st+\sum_{i=1}^Nr_i\Lambda_{bi}({1\over2}\Lambda_{b'i}-u),\;b\neq b'
\end{equation}

\begin{equation}
                                    \label{1.34}
0={\alpha^2\over
4}-su+\sum_{i=1}^Nr_i\Lambda_{bi}({1\over2}\Delta_{ai}-u)
\end{equation}

 From (\ref{1.32}) it follows that
 \begin{equation}
                                    \label{1.35}
v_b^2=v^2=\sigma
\end{equation}

The condition (\ref{1.33}) gives
 \begin{equation}
                                    \label{1.36}
(D-2)({\alpha^2\over
2}+s)-d^2+(D-2)\sum_{i=1}^Nr_i\Lambda_{bi}\Lambda_{b'i}=0,
\end{equation}
that is nothing but (\ref{9}). The relation (\ref{1.34}) coincides
with (\ref{1.31}).

Substituting  (\ref{1.22})--(\ref{1.25}) in (\ref{1.3}) one can check
that equation (\ref{1.3}) is fulfilled. Indeed, a contribution of terms
containing $C_{a}$ in the LHS of (\ref{1.3})
is

$$qt\sum_{a=1}^Eh_a^2C_a+\sum_{i=1}^Nr_i\bigl(\sum_{a=1}^E{1\over
 2}h_a^2C_a\Delta_{ai}-u\sum_{a=1}^Eh_a^2C_a\bigr)-su\sum_{a=1}^Eh_a^2C_a=$$
  $$=\bigl(\sum_{a=1}^Eh_a^2C_a\bigr)(qt-u\sum_{i=1}^N r_i-su)+
\sum_{a=1}^E\sum_{i=1}^N{1\over 2}r_ih_a^2C_a\Delta_{ai}=$$
$$=\bigl(\sum_{a=1}^Eh_a^2C_a\bigr)(q{1\over 2}{D-2-d\over
   D-2}-(D-2-q){1\over 2}{d\over D-2}+{1\over 2}(d-q))=0,$$
i.e. it is zero.
The same is true for the terms containing $\chi _{b}$ ,

 $$qu\sum_{b=1}^Mv_b^2\chi_b-\sum_{i=1}^Nr_i\bigl(\sum_{b=1}^M{1\over
 2}v_b^2\chi_b\Lambda_{bi}-u\sum_{b=1}^Mv_b^2\chi_b\bigr)-
st\sum_{b=1}^Mv_b^2\chi_b=$$
  $$=\bigl(\sum_{b=1}^Mv_b^2\chi_b\bigr)(qu+u\sum_{i=1}^N r_i-st)
-\sum_{b=1}^M\sum_{i=1}^N{1\over 2}r_iv_b^2\chi_b\Lambda_{bi}=$$
$$=\bigl(\sum_{b=1}^Mv_b^2\chi_b\bigr)(q{1\over 2}{d\over
   D-2}+(D-2-q-s){1\over 2}{d\over D-2}-s{D-2-d\over 2(D-2)}-{1\over
 2}(d-s))=0.$$

Non-diagonal part of the Einstein equations gives five types of relations.
One type follows from the requirement of compensation
$\partial_\alpha C_a\partial_\beta C_a$. It has the form
$$[-qt^2-\sum_{i=1}^Nr_i({1\over
4}\Delta_{ai}+u^2-u\Delta_{ai})-
su^2]h_a^4={\alpha^2\over 8}h_a^4-{h_a^2\over 2},$$
that is satisfied due to the requirement 4) and relation (\ref{1.29}).

 Another
type of relations follows from the requirement of compensation
$\partial_\alpha C_a\partial_\beta C_{a'},$ $\,a\not=~a'$
$${\alpha^2\over 8}+qt^2+\sum_{i=1}^Nr_i({1\over
4}\Delta_{ai}\Delta_{a'i}+u^2-u\Delta_{ai})+su^2=0,$$
that holds due to (\ref{7}).

One more type is obtained from the requirement on the coefficient in front of
$\partial_\alpha \chi_b\partial_\beta \chi_b$.  It has the form
$$[-qu^2-\sum_{i=1}^Nr_i({1\over
4}\Lambda_{bi}+u^2-u\Lambda_{bi})-st^2]v_b^4=
{\alpha^2\over 8}v_b^4-{v_b^2\over 2},$$
and takes place due to 4') and (\ref{1.35})

 The next type of relations we get  from the
 requirement of a compensation $\partial_\alpha \chi_b\partial_\beta
 \chi_b',$ $\;b\not=~b'$ $${\alpha^2\over
8}+qu^2+\sum_{i=1}^Nr_i({1\over 2}\Lambda_{bi}-u)({1\over
2}\Lambda_{b'i}-u)+st^2=0,$$ that is equivalent to (\ref{9}).

 The last type of relations follows from the requirement of compensation
 $\partial_\alpha \chi_b\partial_\beta C_a$
and it  looks like
\begin{equation}
                    \label{cc}
{\alpha^2\over 8}-qut+\sum_{i=1}^Nr_i({1\over
2}\Lambda_{bi}-u)({1\over 2}\Delta_{ai}-u)-stu=0.
\end{equation}
Since the LHS of  the above relations is
$${\alpha^2\over
2}(D-2)-d^2+(D-2)\sum_{i=1}^Nr_i\Lambda_{bi}\Delta_{ai}$$
so the requirement (\ref{cc}) is nothing but (\ref{10}).

The form of the metric for this solution is
$$ds^{2}=
(H^{-1}_{1}H^{-1}_{2}...H^{-1}_{E}U_{1}U_{2}...U_{M})^{2t\sigma}
(H_{1}H_{2}...H_{E})^{\sigma}\times $$

$$
\{ (H_{1}H_{2}...H_{E}U_{1}U_{2}...U_{M})^{-\sigma}
\sum _{\mu, \nu =0}^{q-1} \eta_{\mu \nu}dy^{\mu} dy^{\nu}+$$
\begin{equation}
 (U_{1}U_{2}...U_{M})^{-\sigma}\sum _{i}(\prod
_{a}H_{a}^{-\Delta _{ai}} \prod _{b}U_{b}^{\Lambda _{bi}})^{\sigma}\sum
_{m_{i}=1}^{r_{i}}dz_{i}^{m_{i}}dz_{i}^{m_{i}}
+\sum _{\alpha
}dx^{\alpha}dx^{\alpha}\},
                                    \label{8'}
\end{equation}
where harmonic functions $H_a$ and $U_b$ are given by
\begin{equation}
 H_a=1+\sum_{c}{Q_{ac}\over |x-x_{ac}|^s},
\end{equation}
\begin{equation}
 U_b=1+\sum_{c}{P_{bc}\over |x-x_{bc}|^s}.
\end{equation}

This representation of the metric is convenient for calculating the
entropy.  The hole area of the horizon is non-trivial and points
$x_{ac}$ and $x_{bc}$ are not singularities  when
\par
 1)all harmonic functions have the equal number of centers $k$ and
$x_{1c}=\ldots=x_{kc}=x_{c}$.
\par
 2) $2\sigma s(Eu+tM)=2$
\par
3) $\sigma s(\sum_a\Delta_{ai}-\sum_b\Lambda_{bi}+M)=2$
\par
4)$q=1$ or
\par
4')$q=2$ and $s\sigma(E+M)-2=s$
\par
Under these conditions for the case $q=1$ the area of horizon has the
form
\begin{equation}
{\cal
A}_{D-2}=\omega_{s+1}L^{D-s-3}\sum_{c}(\prod_aQ_{ac}\prod_bP_{bc})
^{\sigma\over 2}, \end{equation}

where $\omega_{s+1}$ is volume of $s+1$-dimensional sphere,
\begin{equation}
 \omega_{s+1}={2\pi^{s+2\over 2}\over \Gamma\left({s+2\over 2}\right)}
\end{equation}
and $L$ is a period of all $y_n$, $n=1,..,D-s-3$.
To get non-trivial entropy for $q=2$ one have to make a 'boost'
$$-dy_0^2+dy_1^2\;\longrightarrow\;dudv+K(x)du^2,$$
where
$$u=y_1+y_0,\;v=y_1-y_0,\;K(x)=1+\sum_c{Q_c\over
|x-x_c|^{s}},\;x_c=x_{ac}$$

In this case the area of horizon is given by
\begin{equation}
{\cal A}_{D-2}=\omega_{s+1}L^{D-s-3}\sum_{c}(\prod_aQ_{ac}\prod_bP_{bc})
 ^{\sigma\over 2}Q_c^{1\over 2}
\end{equation}

%%%%%%     PART2.tex

\newpage

\subsection{Examples}

To describe particular examples of the general solution (\ref{8})
it is convenient to present the incidence matrices
$\Delta_{ai}$ and $\Lambda_{bi}$
 by special tables with $E$ (or $M$) rows and $N+2$ columns
 (additional two columns correspond to subspaces of ${\bf M}$ with
dimensions
 $q$ and $s$, briefly $q$-column and $s$-column). If $\Delta_{ai}=1$ we put
 $r_i$ symbols '$\circ$' in an appropriate cell of the table.  If
 $\Lambda_{bi}=1$ we draw $r_i$ symbols '$\bullet$' in a
 corresponding place  of the table. Also we  draw '$\circ$' and
'$\bullet$' in $q$-columns and $s$-columns, respectively. We will not consider
here the cases with  $E+M=1$.

\subsubsection {$D=11$, $d$-form   and $\alpha =0$}
 First we will investigate the case $D=11$ and $\alpha
=0$

 1) {\it ''electric'' ansatz.}

In this case we have the following condition
  \begin{equation}
 \label{2.e}
 q-{d^2\over
 9}+\sum_{i=1}^Nr_i\Delta_{ai}\Delta_{a'i}=0,~~\;a\not= a'
\end{equation}
 where all values are integer.
Therefore there are two types of solutions.
 $$~$$

%\input{pic1_1.pic}
%%%%%%%%%%%%%%%%%%%%%
%%%%%%%%%%%%%%%%%%     PIC1_1.PIC
\unitlength 1.00mm
\linethickness{0.4pt}
\begin{picture}(110.00,15.00)(0.00,25.00)
\put(65.00,25.00){\line(0,1){15.00}}
\put(65.00,40.00){\line(1,0){5.33}}
\put(70.33,40.00){\line(-1,0){5.33}}
\put(65.00,40.00){\line(1,0){5.33}}
\put(70.33,40.00){\line(-1,0){5.33}}
\put(70.00,40.00){\line(0,-1){15.00}}
\put(67.67,37.33){\circle{2.11}}
\put(67.67,32.67){\circle{2.11}}
\put(67.67,27.33){\circle{2.11}}
\put(80.00,40.00){\line(0,-1){15.00}}
\put(90.00,40.00){\line(0,-1){15.00}}
\put(100.00,40.00){\line(0,-1){15.00}}
\put(77.33,37.33){\circle{2.11}}
\put(82.67,32.67){\circle{2.11}}
\put(87.00,32.67){\circle{2.11}}
\put(93.00,27.33){\circle{2.11}}
\put(97.00,27.33){\circle{2.11}}
\put(72.67,37.33){\circle{2.11}}
\put(29.67,32.67){\makebox(0,0)[cc]{d=3, q=1, s=2}}
\put(70.00,40.00){\line(1,0){40.00}}
\put(110.00,40.00){\line(0,-1){15.00}}
\put(110.00,25.00){\line(-1,0){45.00}}
\put(65.00,35.00){\line(1,0){45.00}}
\put(65.00,30.00){\line(1,0){45.00}}
\end{picture}
%%%%%%%%%%%%%%%%%%%%%%%%
%%%%%%%%%%%%%%%%%%%%%%%%%      END OF 1_1

$$ds^2=(H_1H_2H_3)^{1\over3}[-(H_1H_2H_3)^{-1}dy^2+
H_1^{-1}(dy_1^2+dy_2^2)+H_2^{-1}(dy_3^2+dy_4^2)+$$
\begin{equation}
\label{D11e1}
+H_3^{-1}(dy_5^2+dy_6^2)+dx^{\alpha}dx^{\alpha}]
\end{equation}

For $H_{a}$
 in the form
\begin{equation}
 \label{2.2}
H_{a}=1+\sum _{c}\frac{Q_{ac}}{|x-x_{ac}|}
\end{equation}

we get non-zero area of the horizon if all positions of the poles
$x_{ac}$ coincide, $x_{ac}=$$x_{c},$

\begin{equation}
\label{D11e1A}
{\cal A}_9=\omega_3L^6\sum_{c}(Q_{1c}Q_{2c}Q_{3c})^{1/2}
\end{equation}

%\input{pic1_2.pic}
%%%%%%%%%%%%%%%%%%%%%%%%%%%%
%%%%%%%%%%%%%%%%%%%%%%%%%%%%%%     PIC1_2.PIC

\unitlength 1.00mm
\linethickness{0.4pt}
\begin{picture}(109.67,15.00)(0.00,45.00)
\put(64.67,40.00){\line(0,1){15.00}}
\put(74.67,55.00){\line(0,-1){15.00}}
\put(84.67,55.00){\line(0,-1){15.00}}
\put(94.67,55.00){\line(0,-1){15.00}}
\put(87.34,52.67){\circle{2.11}}
\put(92.00,52.67){\circle{2.11}}
\put(67.34,47.67){\circle{2.11}}
\put(72.00,47.67){\circle{2.11}}
\put(87.34,42.67){\circle{2.11}}
\put(92.00,42.67){\circle{2.11}}
\put(77.00,52.67){\circle{2.11}}
\put(82.00,52.67){\circle{2.11}}
\put(77.67,47.67){\circle{2.11}}
\put(82.34,47.67){\circle{2.11}}
\put(97.34,47.67){\circle{2.11}}
\put(102.00,47.67){\circle{2.11}}
\put(97.34,42.67){\circle{2.11}}
\put(102.00,42.67){\circle{2.11}}
\put(67.34,52.67){\circle{2.11}}
\put(72.00,52.67){\circle{2.11}}
\put(67.34,42.67){\circle{2.11}}
\put(72.00,42.67){\circle{2.11}}
\put(29.67,47.67){\makebox(0,0)[cc]{d=6, q=2, s=1}}
\put(64.67,55.00){\line(1,0){45.00}}
\put(109.67,55.00){\line(0,-1){15.00}}
\put(109.67,40.00){\line(-1,0){45.00}}
\put(64.67,40.00){\line(0,1){0.00}}
\put(64.67,50.00){\line(1,0){45.00}}
\put(65.00,45.00){\line(1,0){44.67}}
\end{picture}
%%%%%%%%%%%%%%%%%%%%%%%%%%%%%%        END OF 1_2

$$ds^2=(H_1H_2H_3)^{2\over3}
[(H_1H_2H_3)^{-1}(-dy^2+dy_1^2+K(x)du^2)+(H_1H_2)^{-1}(dy_2^2+dy_3^2)+
$$
\begin{equation}
\label{D11e2}
(H_1H_3)^{-1}(
dy_4^2+dy_5^2)+(H_2H_3)^{-1}(dy_6^2+dy_7^2)+dx^{\alpha}dx^{\alpha}]
\end{equation}

\begin{equation}
\label{D11e2A}
{\cal A}_9=4\pi L^7\sum_{c}(Q_cQ_{1c}Q_{2c}Q_{3c})^{1/2}
\end{equation}

Note that the solutions presented in these tables correspond to
"maximal" configuration in the sense that apart from the ansatzes
presented by the above  tables one can also consider ansatzes
which can be obtained
 by deleting some rows of our matrices, but these solutions have zero area
of the horizon.

The solution (\ref{D11e1}) has been discussed in \cite{Ts1,IN},
the solution (\ref{D11e2}) was considered
in ~\cite{klebanov}.
\par
  2) {\it ''magnetic'' ansatz}.

 In this case we have the following condition
  \begin{equation}
 \label{2.m}
s-{d^2\over
 9}+\sum_{i=1}^Nr_i\Lambda_{bi}\Lambda_{b'i}=0,\;b\not= b',
\end{equation}
where all values are integer. Therefore there are similary two types
 of solutions.  $$~$$
%\input{pic2_1.pic}
%%%%%%%%%%%%%%%%%%%%        PIC2_1.PIC
\unitlength 1.00mm
\linethickness{0.4pt}
\begin{picture}(110.33,10.67)(0.00,10.00)
\put(65.33,7.67){\line(0,1){15.00}}
\put(75.33,22.67){\line(0,-1){15.00}}
\put(85.33,22.67){\line(0,-1){15.00}}
\put(95.33,22.67){\line(0,-1){15.00}}
\put(105.33,22.67){\line(0,-1){15.00}}
\put(30.33,15.34){\makebox(0,0)[cc]{d=3, q=2, s=1}}
\put(108.00,20.34){\circle*{2.11}}
\put(108.00,15.34){\circle*{2.11}}
\put(108.00,10.34){\circle*{2.11}}
\put(98.00,20.34){\circle*{2.11}}
\put(103.00,20.34){\circle*{2.11}}
\put(87.66,15.34){\circle*{2.11}}
\put(93.00,15.34){\circle*{2.11}}
\put(78.00,10.34){\circle*{2.11}}
\put(83.00,10.34){\circle*{2.11}}
\put(65.33,22.67){\line(1,0){45.00}}
\put(110.33,22.67){\line(0,-1){15.00}}
\put(110.33,7.67){\line(-1,0){45.00}}
\put(65.33,17.67){\line(1,0){45.00}}
\put(65.33,12.67){\line(1,0){45.00}}
\end{picture}
%%%%%%%%%%%%%%%%%%%%%%%%%%%%%%%%        END OF 2_1

$$ds^2=(U_1U_2U_3)^{2\over3}[(U_1U_2U_3)^{-1}(-dy^2+dy_1^2+K(x)du^2)+
(U_1U_2)^{-1}
(dy_2^2+dy_3^2)+$$
\begin{equation}
\label{D11m1}
(U_1U_3)^{-1}(dy_4^2+dy_5^2)+
(U_2U_3)^{-1}(dy_6^2+dy_7^2)+dx^{\alpha}dx^{\alpha}]
\end{equation}

\begin{equation}
\label{D11m1A}
{\cal A}_9=4\pi L^7\sum_{c}(Q_cP_{1c}P_{2c}P_{3c})^{1/2}
\end{equation}

%\input{pic2_2.pic}
%%%%%%%%%%%%%%%%%%%%%%%%       PIC2_2.PIC

\unitlength 1.00mm
\linethickness{0.4pt}
\begin{picture}(110.00,23.00)(0.00,5.00)
\put(65.00,8.00){\line(0,1){15.00}}
\put(65.00,23.00){\line(1,0){5.33}}
\put(70.33,23.00){\line(-1,0){5.33}}
\put(65.00,23.00){\line(1,0){5.33}}
\put(70.33,23.00){\line(-1,0){5.33}}
\put(70.00,23.00){\line(0,-1){15.00}}
\put(80.00,23.00){\line(0,-1){15.00}}
\put(90.00,23.00){\line(0,-1){15.00}}
\put(100.00,23.00){\line(0,-1){15.00}}
\put(29.67,15.67){\makebox(0,0)[cc]{d=6, q=1, s=2}}
\put(102.67,20.67){\circle*{2.11}}
\put(107.67,20.67){\circle*{2.11}}
\put(92.67,20.67){\circle*{2.11}}
\put(97.33,20.67){\circle*{2.11}}
\put(102.67,15.67){\circle*{2.11}}
\put(107.67,15.67){\circle*{2.11}}
\put(92.67,15.67){\circle*{2.11}}
\put(97.33,15.67){\circle*{2.11}}
\put(102.67,10.67){\circle*{2.11}}
\put(107.67,10.67){\circle*{2.11}}
\put(72.67,15.67){\circle*{2.11}}
\put(77.67,15.67){\circle*{2.11}}
\put(72.67,10.67){\circle*{2.11}}
\put(77.67,10.67){\circle*{2.11}}
\put(82.67,20.67){\circle*{2.11}}
\put(87.33,20.67){\circle*{2.11}}
\put(82.67,10.67){\circle*{2.11}}
\put(87.33,10.67){\circle*{2.11}}
\put(70.00,23.00){\line(1,0){40.00}}
\put(110.00,23.00){\line(0,-1){15.00}}
\put(110.00,8.00){\line(-1,0){45.00}}
\put(65.00,18.00){\line(1,0){45.00}}
\put(65.00,13.00){\line(1,0){45.00}}
\end{picture}
%%%%%%%%%%%%%%%%%%%%%%%%%%%%%%          END OF 2_2

$$ds^2=(U_1U_2U_3)^{1\over3}[-(U_1U_2U_3)^{-1}dy^2+U_1^{-1}(dy_1^2+dy_2^2)
+U_2^{-1}(dy_3^2+dy_4^2)+$$
\begin{equation}
\label{D11m2}
+U_3^{-1}(dy_5^2+dy_6^2)+dx^{\alpha}dx^{\alpha}]
\end{equation}
\begin{equation}
\label{D11m2A}
{\cal A}_9=\omega_3L^6\sum_{c}(P_{1c}P_{2c}P_{3c})^{1/2}
\end{equation}

These solutions are dual to the 'electric' ones.

3){\it ''electric'' $+$ ''magnetic'' ansatz.}

In this case we have the additional condition
 \begin{equation}
 \label{2.em}
-{d^2\over 9}+\sum_{i=1}^Nr_i\Lambda_{bi}\Delta_{ai}=0
\end{equation}
 Our strategy in finding solutions of
(\ref{2.e}), (\ref{2.m}) and (\ref{2.em}) is very simple. We take
tables corresponding to electric and magnetic ansatzes and try to put
on the corresponding configurations of the fields the additional
condition (\ref{2.em}). One can check that we cannot combine the
maximal ansatz presented by (\ref{D11e1}) (or (\ref{D11e2})) with
maximal magnetic ansatz corresponding to (\ref{D11m1})  (or
(\ref{D11m2})), but we can combine a part of table (\ref{D11e1}) with
a part of table (\ref{D11m1}) as well as   a part of table (\ref{D11e2}) with a
part of table (\ref{D11m2})  as it is presented in the tables connected
with solutions (\ref{D11em1}) and (\ref{D11em2}),
respectively.  So,  we have two types of solutions.  $$~$$

$$~$$
%\input{pic3_1.pic}
%%%%%%%%%%%%%%%%%%%%%%%%%%%      PIC3_1.pic

\unitlength 1.00mm
\linethickness{0.4pt}
\begin{picture}(110.00,10.00)(0.00,5.00)
\put(65.00,13.00){\line(0,1){15.00}}
\put(65.00,28.00){\line(1,0){5.33}}
\put(70.33,28.00){\line(-1,0){5.33}}
\put(65.00,28.00){\line(1,0){5.33}}
\put(70.33,28.00){\line(-1,0){5.33}}
\put(70.00,28.00){\line(0,-1){15.00}}
\put(67.67,25.33){\circle{2.00}}
\put(67.67,20.67){\circle{2.00}}
\put(80.00,28.00){\line(0,-1){15.00}}
\put(90.00,28.00){\line(0,-1){15.00}}
\put(77.33,25.33){\circle{2.11}}
\put(82.67,20.67){\circle{2.11}}
\put(87.00,20.67){\circle{2.11}}
\put(72.67,25.33){\circle{2.11}}
\put(29.67,20.67){\makebox(0,0)[cc]{d=3, q=1, s=1}}
\put(65.00,13.00){\line(0,-1){5.00}}
\put(70.00,13.00){\line(0,-1){5.00}}
\put(80.00,13.00){\line(0,-1){5.00}}
\put(90.00,13.00){\line(0,-1){5.00}}
\put(105.00,28.00){\line(0,-1){20.00}}
\put(107.67,15.67){\circle*{2.11}}
\put(107.67,10.67){\circle*{2.11}}
\put(72.67,15.67){\circle*{2.11}}
\put(82.67,15.67){\circle*{2.11}}
\put(77.33,10.67){\circle*{2.11}}
\put(87.00,10.67){\circle*{2.11}}
\put(70.00,28.00){\line(1,0){40.00}}
\put(110.00,28.00){\line(0,-1){20.00}}
\put(110.00,8.00){\line(-1,0){45.00}}
\put(65.00,23.00){\line(1,0){45.00}}
\put(65.00,18.00){\line(1,0){45.00}}
\put(65.00,13.00){\line(1,0){45.00}}
\put(75.00,28.00){\line(0,-1){20.00}}
\put(85.00,28.00){\line(0,-1){20.00}}
\end{picture}

%%%%%%%%%%%%%%%%%%%%%%%%%%%%%%%%%%%%%%      END OF 3_1
$$ds^2=(H_1H_2)^{1\over3}(U_1U_2)^
{2\over3}[-(H_1H_2U_1U_2)^{-1}dy^2+(H_1U_2)^{-1}dy_1^2+(H_1U_1)^{-1}dy_2^2+$$
\begin{equation}
\label{D11em1}
+(H_2U_2)^{-1}dy_3^2+{H_2U_1}^{-1}dy_4^2
+(U_1U_2)^{-1}(dy_5^2+dy_6^2+dy_7^2)+dx^{\alpha}dx^{\alpha}]
\end{equation}

\begin{equation}
\label{D11em1A}
{\cal A}_9=4\pi L^7\sum_{c}(Q_{1c}Q_{2c}P_{1c}P_{2c})^{1/2}
\end{equation}

%\input{pic3_2.pic}
%%%%%%%%%%%%%%%%%%%%%%%%%%%      PIC3_2.PIC
\unitlength 1.00mm
\linethickness{0.4pt}
\begin{picture}(109.67,20.67)(0.00,10.00)
\put(64.67,12.67){\line(0,1){15.00}}
\put(84.67,27.67){\line(0,-1){15.00}}
\put(94.67,27.67){\line(0,-1){15.00}}
\put(104.67,27.67){\line(0,-1){15.00}}
\put(87.34,25.34){\circle{2.11}}
\put(92.00,25.34){\circle{2.11}}
\put(67.34,20.34){\circle{2.11}}
\put(72.00,20.34){\circle{2.11}}
\put(77.00,25.34){\circle{2.11}}
\put(82.00,25.34){\circle{2.11}}
\put(77.67,20.34){\circle{2.11}}
\put(82.34,20.34){\circle{2.11}}
\put(97.34,20.34){\circle{2.11}}
\put(102.00,20.34){\circle{2.11}}
\put(67.34,25.34){\circle{2.11}}
\put(72.00,25.34){\circle{2.11}}
\put(29.67,20.34){\makebox(0,0)[cc]{d=6, q=1, s=1}}
\put(64.67,12.67){\line(0,-1){5.00}}
\put(104.67,12.67){\line(0,-1){5.00}}
\put(94.67,12.67){\line(0,-1){5.00}}
\put(84.67,12.67){\line(0,-1){5.00}}
\put(69.67,27.67){\line(0,-1){20.00}}
\put(69.67,27.67){\line(1,0){15.00}}
\put(74.67,27.67){\line(0,1){0.00}}
\put(74.67,27.67){\line(0,1){0.00}}
\put(69.67,27.67){\line(1,0){15.00}}
\put(107.34,15.34){\circle*{2.11}}
\put(107.34,10.34){\circle*{2.11}}
\put(84.67,7.67){\line(0,1){5.00}}
\put(84.67,7.67){\line(0,1){5.00}}
\put(72.00,15.34){\circle*{2.11}}
\put(77.00,15.34){\circle*{2.11}}
\put(81.34,15.34){\circle*{2.11}}
\put(72.00,10.34){\circle*{2.11}}
\put(77.00,10.34){\circle*{2.11}}
\put(81.34,10.34){\circle*{2.11}}
\put(87.34,15.00){\circle*{2.11}}
\put(92.00,10.34){\circle*{2.11}}
\put(97.34,15.00){\circle*{2.11}}
\put(102.00,10.34){\circle*{2.11}}
\put(64.67,27.67){\line(1,0){5.00}}
\put(84.67,27.67){\line(1,0){25.00}}
\put(109.67,27.67){\line(0,-1){20.00}}
\put(109.67,7.67){\line(-1,0){45.00}}
\put(64.67,22.67){\line(1,0){45.00}}
\put(64.67,17.67){\line(1,0){45.00}}
\put(64.67,12.67){\line(1,0){45.00}}
\put(90.00,27.67){\line(0,-1){20.00}}
\put(100.00,27.67){\line(0,-1){20.00}}
\end{picture}
%%%%%%%%%%%%%%%%%%%%%%%%%%%%%         END OF 3_2

$$ds^2=(H_1H_2)^{2\over3}(U_1U_2)^{1
\over3}[-(H_1H_2U_1U_2)^{-1}dy^2+(H_1H_2)^{-1}(dy_1^2+dy_2^2+dy_3^2)
+(H_1U_2)^{-1}dy_4^2+$$
\begin{equation}
\label{D11em2}
+(H_1U_1)^{-1}dy_5^2+(H_2U_2)^{-1}dy_6^2
+(H_2U_1)^{-1}dy_7^2+dx^{\alpha}dx^{\alpha}]
\end{equation}

\begin{equation}
\label{D11em2A}
{\cal A}_9=4\pi L^7\sum_{c}(Q_{1c}Q_{2c}P_{1c}P_{2c})^{1/2}
\end{equation}

 The solution (\ref{D11em1}) has been constructed in ~\cite{klebanov} and it
is dual to (\ref{D11em2}).
\subsubsection{D=10, d-form and $\alpha \neq 0$.}

1){\it ''electric'' ansatz.}

 In this case we have the following condition
 $$ {\alpha^2\over 2}+q-{d^2\over
 9}+\sum_{i=1}^Nr_i\Delta_{ai}\Delta_{a'i}=0,\;a\not= a',$$
 where all values are integer. Therefore there are four types of solutions.
 $$~$$

%\input{pic7_1.pic}
%%%%%%%%%%%%%%%%%%%%%%%%%%%%%%%%          PIC7_1.PIC
\special{em:linewidth 0.4pt}
\unitlength 1.00mm
\linethickness{0.4pt}
\begin{picture}(110.00,10.00)(0.00,10.00)
\emline{70.00}{8.00}{1}{70.00}{23.00}{2}
\emline{75.00}{23.00}{3}{75.00}{8.00}{4}
\emline{85.00}{23.00}{5}{85.00}{8.00}{6}
\emline{95.00}{23.00}{7}{95.00}{8.00}{8}
\emline{105.00}{23.00}{9}{105.00}{8.00}{10}
\put(72.67,20.67){\circle{2.11}}
\put(77.67,20.67){\circle{2.11}}
\put(82.33,20.67){\circle{2.11}}
\put(72.67,15.67){\circle{2.11}}
\put(87.67,15.67){\circle{2.11}}
\put(92.67,15.67){\circle{2.11}}
\put(72.67,10.67){\circle{2.11}}
\put(98.00,10.67){\circle{2.11}}
\put(102.67,10.67){\circle{2.11}}
\put(40.00,15.67){\makebox(0,0)[cc]{d=3, $\alpha=\pm {1\over 2}$, q=1, s=1}}
\emline{70.00}{23.00}{11}{110.00}{23.00}{12}
\emline{110.00}{23.00}{13}{110.00}{8.00}{14}
\emline{110.00}{8.00}{15}{70.00}{8.00}{16}
\emline{70.00}{18.00}{17}{110.00}{18.00}{18}
\emline{70.00}{13.00}{19}{110.00}{13.00}{20}
\end{picture}

%%%%%%%%%%%%%%%%%%%%%%%%%%%%%%%%%%%%        END OF 7_1
$$ds^2=(H_1H_2H_3)^{3\over8}[-(
H_1H_2H_3)^{-1}dy^2+H_1^{-1}(dy_1^2+dy_2^2)+H_2^{-1}(dy_3^2+dy_4^2)+$$
\begin{equation}
\label{D10e1}
+H_3^{-1}(dy_5^2+dy_6^2)+dx^{\alpha}dx^{\alpha}]
\end{equation}

All centers of harmonic functions are points of singularity.
$$~$$

%\input{pic7_2.pic}

%%%%%%%%%%%%%%%%%%%%%%%%%%%%%%%%%%%%%      PIC7_2.PIC
\special{em:linewidth 0.4pt}
\unitlength 1.00mm
\linethickness{0.4pt}
\begin{picture}(110.00,18.00)(0.00,5.00)
\emline{70.00}{18.00}{1}{70.00}{8.00}{2}
\emline{75.00}{18.00}{3}{75.00}{8.00}{4}
\emline{90.00}{18.00}{5}{90.00}{8.00}{6}
\emline{105.00}{18.00}{7}{105.00}{8.00}{8}
\put(72.67,15.33){\circle{2.11}}
\put(77.67,15.33){\circle{2.11}}
\put(82.33,15.33){\circle{2.11}}
\put(87.67,15.33){\circle{2.11}}
\put(72.67,10.67){\circle{2.11}}
\put(92.67,10.67){\circle{2.11}}
\put(98.00,10.67){\circle{2.11}}
\put(102.67,10.67){\circle{2.11}}
\put(40.00,13.00){\makebox(0,0)[cc]{d=4,$\alpha=\pm\sqrt{2}$, q=1, s=1}}
\emline{70.00}{18.00}{9}{110.00}{18.00}{10}
\emline{110.00}{18.00}{11}{110.00}{8.00}{12}
\emline{110.00}{8.00}{13}{70.00}{8.00}{14}
\emline{70.00}{13.00}{15}{110.00}{13.00}{16}
\end{picture}

%%%%%%%%%%%%%%%%%%%%%%%%     END OF 7_2

$$ds^2=(H_1H_2)^{1\over3}[-(H_1H_2)^{-{2\over3}}dy^2
+H_1^{-{2\over3}}(dy_1^2+dy_2^2+dy_3^2)+$$
\begin{equation}
\label{D10e2}
+H_2^{-{2\over3}}(dy_4^2+dy_5^2+dy_6^2)+dx^{\alpha}dx^{\alpha}]
\end{equation}
The area of horizon is equal to zero.

%\input{pic7_3.pic}
%%%%%%%%%%%%%%%%%%%%%%%%%%%%%%%%%  PIC7_3.PIC
\special{em:linewidth 0.4pt}
\unitlength 1.00mm
\linethickness{0.4pt}
\begin{picture}(110.00,20.00)(0.00,10.00)
\emline{70.00}{28.00}{1}{70.00}{8.00}{2}
\emline{75.00}{28.00}{3}{75.00}{8.00}{4}
\put(72.67,25.67){\circle{2.11}}
\put(77.67,25.67){\circle{2.11}}
\put(82.33,25.67){\circle{2.11}}
\put(87.67,25.67){\circle{2.11}}
\put(72.67,20.67){\circle{2.11}}
\put(77.67,20.67){\circle{2.11}}
\put(92.67,20.67){\circle{2.11}}
\put(98.00,20.67){\circle{2.11}}
\emline{105.00}{28.00}{5}{105.00}{8.00}{6}
\put(72.67,15.67){\circle{2.11}}
\put(82.33,15.67){\circle{2.11}}
\put(92.67,15.67){\circle{2.11}}
\put(102.67,15.67){\circle{2.11}}
\put(72.67,10.67){\circle{2.11}}
\put(87.67,10.67){\circle{2.11}}
\put(98.00,10.67){\circle{2.11}}
\put(102.67,10.67){\circle{2.11}}
\emline{90.00}{28.00}{7}{90.00}{8.00}{8}
\put(40.00,18.00){\makebox(0,0)[cc]{d=4, $\alpha=0$, q=1, s=1}}
\emline{70.00}{28.00}{9}{110.00}{28.00}{10}
\emline{110.00}{28.00}{11}{110.00}{8.00}{12}
\emline{110.00}{8.00}{13}{70.00}{8.00}{14}
\emline{70.00}{23.00}{15}{110.00}{23.00}{16}
\emline{70.33}{18.00}{17}{110.00}{18.00}{18}
\emline{70.00}{13.00}{19}{110.00}{13.00}{20}
\emline{80.00}{28.00}{21}{80.00}{8.00}{22}
\emline{85.00}{28.00}{23}{85.00}{8.00}{24}
\emline{95.00}{28.00}{25}{95.00}{8.00}{26}
\emline{100.00}{28.00}{27}{100.00}{8.00}{28}
\end{picture}

%%%%%%%%%%%%%%%%%%%%%%%%%%%%%%%%%%%   END OF 7_3

$$ds^2=(H_1H_2H_3H_4)^{1\over2}[-(H_1H_2H_3H_4)^{-1}dy^2+
(H_1H_2)^{-1}dy_1^2+(H_1H_3)^{-1}dy_2^2+(H_1H_4)^{-1}dy_3^2+$$
\begin{equation}
\label{D10e3}
+(H_2H_3)^{-1}dy_4^2+(H_2H_4)^{-1}dy_5^2
+(H_3H_4)^{-1}dy_6^2+dx^{\alpha}dx^{\alpha}]
\end{equation}

\begin{equation}
{\cal A}_8=4\pi L^6
\sum_{c}(Q_{1c}Q_{2c}Q_{3c}Q_{4c})^{1/2}
\end{equation}

This solution was discussed in ~\cite{Berg}.

%\input{pic7_4.pic}
%%%%%%%%%%%%%%%%%%%%%%%%%%%%%%%%%%     PIC7_4.PIC
\special{em:linewidth 0.4pt}
\unitlength 1.00mm
\linethickness{0.4pt}
\begin{picture}(110.00,20.00)(0.00,7.00)
\emline{70.00}{8.00}{1}{70.00}{23.00}{2}
\emline{75.00}{23.00}{3}{75.00}{8.00}{4}
\emline{85.00}{23.00}{5}{85.00}{8.00}{6}
\emline{95.00}{23.00}{7}{95.00}{8.00}{8}
\emline{105.00}{23.00}{9}{105.00}{8.00}{10}
\put(72.67,20.33){\circle{2.11}}
\put(77.67,20.33){\circle{2.11}}
\put(82.33,20.33){\circle{2.11}}
\put(88.00,20.33){\circle{2.11}}
\put(92.67,20.33){\circle{2.11}}
\put(72.67,15.67){\circle{2.11}}
\put(77.67,15.67){\circle{2.11}}
\put(82.33,15.67){\circle{2.11}}
\put(98.00,15.67){\circle{2.11}}
\put(102.67,15.67){\circle{2.11}}
\put(72.67,10.67){\circle{2.11}}
\put(88.00,10.67){\circle{2.11}}
\put(92.67,10.67){\circle{2.11}}
\put(98.00,10.67){\circle{2.11}}
\put(102.67,10.67){\circle{2.11}}
\put(39.67,15.67){\makebox(0,0)[cc]{d=5, $\alpha=\pm{1\over 2}$, q=1,s=1}}
\emline{70.00}{23.00}{11}{110.00}{23.00}{12}
\emline{110.00}{23.00}{13}{110.00}{8.00}{14}
\emline{110.00}{8.00}{15}{70.00}{8.00}{16}
\emline{70.00}{18.00}{17}{110.00}{18.00}{18}
\emline{70.00}{13.00}{19}{110.00}{13.00}{20}
\end{picture}

$$ds^2=(H_1H_2H_3)^{5\over8}
[-(H_1H_2H_3)^{-1}dy^2+(H_1H_2)^{-1}(dy_1^2+dy_2^2)+(H_1H_3)^{-1}(dy_3^2+
dy_4^2)+$$
\begin{equation}
\label{D10e4}
+(H_2H_3)^{-1}(dy_5^2+dy_6^2)+dx^{\alpha}dx^{\alpha}]
\end{equation}

The area of horizon is equal to zero.

%%%%%%%%%%%%%%%%%%%%%%%%%%%%%%%%%%%%%%%%%%%       END of 7_4

  2) {\it ''magnetic'' ansatz.}

In this case we have the following condition
 $$ {\alpha^2\over 2}+s-{d^2\over
 9}+\sum_{i=1}^Nr_i\Lambda_{bi}\Lambda_{b'i}=0\;b\not= b'$$
 where all values are integer. Therefore there are also four  types
 of solutions.
  $$~$$
%\input{pic8_1.pic}
%%%%%%%%%%%%%%%%%%%%%%%%%%%%%%%%%%%%       PIC8_1.pic
\special{em:linewidth 0.4pt}
\unitlength 1.00mm
\linethickness{0.4pt}
\begin{picture}(110.00,15.00)(0.00,5.00)
\emline{70.00}{8.00}{1}{70.00}{23.00}{2}
\emline{75.00}{23.00}{3}{75.00}{8.00}{4}
\emline{85.00}{23.00}{5}{85.00}{8.00}{6}
\emline{95.00}{23.00}{7}{95.00}{8.00}{8}
\emline{105.00}{23.00}{9}{105.00}{8.00}{10}
\put(107.33,20.67){\circle*{2.11}}
\put(97.33,20.67){\circle*{2.11}}
\put(102.33,20.67){\circle*{2.11}}
\put(107.33,15.67){\circle*{2.11}}
\put(87.67,15.67){\circle*{2.11}}
\put(92.67,15.67){\circle*{2.11}}
\put(77.67,10.67){\circle*{2.11}}
\put(82.67,10.67){\circle*{2.11}}
\put(107.33,10.67){\circle*{2.11}}
\put(40.00,15.67){\makebox(0,0)[cc]{d=3, $\alpha=\pm{1\over 2}$, q=1, s=1}}
\emline{70.00}{23.00}{11}{110.00}{23.00}{12}
\emline{110.00}{23.00}{13}{110.00}{8.00}{14}
\emline{110.00}{8.00}{15}{70.00}{8.00}{16}
\emline{70.00}{18.00}{17}{110.00}{18.00}{18}
\emline{70.00}{13.00}{19}{110.00}{13.00}{20}
 \end{picture}

$$ds^2=(U_1U_2U_3)^{5\over8}[-(U_1U_2U_3)^{-1}dy^2+
(U_1U_2)^{-1}(dy_1^2+dy_2^2)+(U_1U_3)^{-1}(dy_3^2+dy_4^2)+$$
\begin{equation}
\label{D10m1}
+(U_2U_3)^{-1}(dy_5^2+dy_6^2)+dx^{\alpha}dx^{\alpha}]
\end{equation}

 The entropy is equal to zero.
%%%%%%%%%%%%%%%%%%%%%%%%%%%%%%%%%%%%%%%    END OF 8_1

%\input{pic8_2.pic}
%%%%%%%%%%%%%%%%%%%%%%%%%%%%%%%           PIC8_2.PIC
\special{em:linewidth 0.4pt}
\unitlength 1.00mm
\linethickness{0.4pt}
\begin{picture}(110.00,12.00)(0.00,10.00)
\emline{70.00}{18.00}{1}{70.00}{8.00}{2}
\emline{75.00}{18.00}{3}{75.00}{8.00}{4}
\emline{90.00}{18.00}{5}{90.00}{8.00}{6}
\emline{105.00}{18.00}{7}{105.00}{8.00}{8}
\put(107.33,15.67){\circle*{2.11}}
\put(107.33,10.67){\circle*{2.11}}
\put(92.67,15.67){\circle*{2.11}}
\put(97.33,15.67){\circle*{2.11}}
\put(102.33,15.67){\circle*{2.11}}
\put(77.67,10.67){\circle*{2.11}}
\put(82.67,10.67){\circle*{2.11}}
\put(87.67,10.67){\circle*{2.11}}
\put(40.00,13.00){\makebox(0,0)[cc]{d=4, $\alpha=\pm\sqrt{2}$, q=1, s=1}}
\emline{70.00}{18.00}{9}{110.00}{18.00}{10}
\emline{110.00}{18.00}{11}{110.00}{8.00}{12}
\emline{110.00}{8.00}{13}{70.00}{8.00}{14}
\emline{70.00}{13.00}{15}{110.00}{13.00}{16}
\end{picture}

$$ds^2=(U_1U_2)^{1\over3}[-(U_1U_2)^{-{2\over3}}dy^2+
U_1^{-{2\over3}}(dy_1^2+dy_2^2+dy_3^2)+$$
\begin{equation}
\label{D10m2}
+U_2^{-{2\over3}}(dy_4^2+dy_5^2+dy_6^2)+dx^{\alpha}dx^{\alpha}]
\end{equation}

The area of horizon is equal to zero.

%%%%%%%%%%%%%%%%%%%%%%%%%%%%%%%            END OF 8_2

%\input{pic8_3.pic}
%%%%%%%%%%%%%%%%%%%%%%%%%%%%%%%    PIC8_3.PIC
\special{em:linewidth 0.4pt}
\unitlength 1.00mm
\linethickness{0.4pt}
\begin{picture}(110.00,20.00)(0.00,10.00)
\emline{70.00}{28.00}{1}{70.00}{8.00}{2}
\emline{75.00}{28.00}{3}{75.00}{8.00}{4}
\emline{105.00}{28.00}{5}{105.00}{8.00}{6}
\emline{105.00}{28.00}{7}{105.00}{18.00}{8}
\put(107.33,25.67){\circle*{2.11}}
\put(107.33,20.67){\circle*{2.11}}
\emline{105.00}{18.00}{9}{105.00}{8.00}{10}
\put(107.33,15.67){\circle*{2.11}}
\put(107.33,10.67){\circle*{2.11}}
\put(92.67,25.67){\circle*{2.11}}
\put(97.33,25.67){\circle*{2.11}}
\put(102.33,25.67){\circle*{2.11}}
\put(102.33,20.67){\circle*{2.11}}
\put(87.67,20.67){\circle*{2.11}}
\put(82.67,20.67){\circle*{2.11}}
\put(97.33,15.67){\circle*{2.11}}
\put(87.67,15.67){\circle*{2.11}}
\put(77.67,15.67){\circle*{2.11}}
\put(92.67,10.67){\circle*{2.11}}
\put(82.67,10.67){\circle*{2.11}}
\put(77.67,10.67){\circle*{2.11}}
\put(40.00,18.00){\makebox(0,0)[cc]{d=4, $\alpha=0$, q=1, s=1}}
\emline{70.00}{28.00}{11}{110.00}{28.00}{12}
\emline{110.00}{28.00}{13}{110.00}{8.00}{14}
\emline{110.00}{8.00}{15}{70.00}{8.00}{16}
\emline{70.00}{23.00}{17}{110.00}{23.00}{18}
\emline{70.00}{18.00}{19}{110.00}{18.00}{20}
\emline{70.00}{13.00}{21}{110.00}{13.00}{22}
\emline{80.00}{28.00}{23}{80.00}{8.00}{24}
\emline{85.00}{28.00}{25}{85.00}{8.00}{26}
\emline{90.00}{28.00}{27}{90.00}{8.00}{28}
\emline{95.00}{28.00}{29}{95.00}{8.00}{30}
\emline{100.00}{28.00}{31}{100.00}{8.00}{32}
\end{picture}

$$ds^2=(U_1U_2U_3U_4)^{1\over2}[-(U_1U_2U_3U_4)^{-1}dy^2+
(U_1U_2)^{-1}dy_1^2+(U_1U_3)^{-1}dy_2^2+(U_1U_4)^{-1}dy_3^2+$$
\begin{equation}
\label{D10m3}
+(U_2U_3)^{-1}dy_4^2+(U_2U_4)^{-1}dy_5^2+(U_3U_4)^{-1}dy_6^2+
dx^{\alpha}dx^{\alpha}]
\end{equation}

\begin{equation}
{\cal A}_8=4\pi L^6
\sum_{c}(Q_{1c}Q_{2c}Q_{3c}Q_{4c})^{1/2}
\end{equation}

%%%%%%%%%%%%%%%%%%%%%%%%%%%%%%%%       END OF 8_3

%\input{pic8_4.pic}
%%%%%%%%%%%%%%%%%%%%%%%%%%%%%%        PIC8_4.PIC
\special{em:linewidth 0.4pt}
\unitlength 1.00mm
\linethickness{0.4pt}
\begin{picture}(110.33,13.67)(0.00,10.00)
\emline{70.33}{7.67}{1}{70.33}{22.67}{2}
\emline{75.33}{22.67}{3}{75.33}{7.67}{4}
\emline{85.33}{22.67}{5}{85.33}{7.67}{6}
\emline{95.33}{22.67}{7}{95.33}{7.67}{8}
\emline{105.33}{22.67}{9}{105.33}{7.67}{10}
\put(107.66,20.34){\circle*{2.11}}
\put(93.00,20.34){\circle*{2.11}}
\put(97.66,20.34){\circle*{2.11}}
\put(102.66,20.34){\circle*{2.11}}
\put(107.66,15.34){\circle*{2.11}}
\put(97.66,15.34){\circle*{2.11}}
\put(102.66,15.34){\circle*{2.11}}
\put(88.00,20.34){\circle*{2.11}}
\put(78.00,15.34){\circle*{2.11}}
\put(83.00,15.34){\circle*{2.11}}
\put(78.00,10.34){\circle*{2.11}}
\put(83.00,10.34){\circle*{2.11}}
\put(88.00,10.34){\circle*{2.11}}
\put(93.00,10.34){\circle*{2.11}}
\put(107.66,10.00){\circle*{2.11}}
\put(40.00,15.34){\makebox(0,0)[cc]{d=5, $\alpha=\pm{1\over 2}$, q=1, s=1}}
\emline{70.33}{22.67}{11}{110.33}{22.67}{12}
\emline{110.33}{22.67}{13}{110.33}{7.67}{14}
\emline{110.33}{7.67}{15}{70.33}{7.67}{16}
\emline{70.33}{17.67}{17}{110.33}{17.67}{18}
\emline{70.33}{12.67}{19}{110.33}{12.67}{20}
\end{picture}

$$ds^2=(U_1U_2U_3)^{3\over8}[-(U_1U_2U_3)^{-1}dy^2+
U_1^{-1}(dy_1^2+dy_2^2)+U_2^{-1}(dy_3^2+dy_4^2)+$$
\begin{equation}
\label{D10m4}
+U_3^{-1}(dy_5^2+dy_6^2)+dx^{\alpha}dx^{\alpha}]
\end{equation}

This metric has singularities in the centers of $U_b$

%%%%%%%%%%%%%%%%%%%%%%%%%%%%%                END OF 8_4

The solutions (\ref{D10m1}),(\ref{D10m2}),(\ref{D10m3})
 and (\ref{D10m4}) are dual to
 (\ref{D10e4}),(\ref{D10e2}),(\ref{D10e3}) and (\ref{D10e1}) respectively.

 3) {\it ''electric'' $+$ ''magnetic'' ansatz.}
In this case we have the additional condition
$${\alpha^2\over 2}-{d^2\over 9}+\sum_{i=1}^Nr_i\Lambda_{bi}\Delta_{ai}=0 $$
Again we have three types of solutions.
$$~$$
%\input{pic9_1.pic}
%%%%%%%%%%%%%%%%%%%%%%%%%%%%%%%%%%%           PIC9_1.PIC
\special{em:linewidth 0.4pt}
\unitlength 1.00mm
\linethickness{0.4pt}
\begin{picture}(110.00,28.00)
\emline{70.00}{8.00}{1}{70.00}{28.00}{2}
\emline{75.00}{28.00}{3}{75.00}{8.00}{4}
\emline{85.00}{28.00}{5}{85.00}{8.00}{6}
\emline{95.00}{28.00}{7}{95.00}{8.00}{8}
\emline{105.00}{28.00}{9}{105.00}{8.00}{10}
\put(72.67,25.33){\circle{2.11}}
\put(77.67,25.33){\circle{2.11}}
\put(82.33,25.33){\circle{2.11}}
\put(72.67,20.67){\circle{2.11}}
\put(87.33,20.67){\circle{2.11}}
\put(92.67,20.67){\circle{2.11}}
\put(107.67,15.67){\circle*{2.11}}
\put(77.67,15.67){\circle*{2.11}}
\put(87.33,15.67){\circle*{2.11}}
\put(82.33,10.67){\circle*{2.11}}
\put(92.67,10.67){\circle*{2.11}}
\put(107.67,10.67){\circle*{2.11}}
\put(40.00,18.00){\makebox(0,0)[cc]{d=3, $\alpha=\pm{1\over 2}$, q=1, s=1}}
\emline{70.00}{28.00}{11}{110.00}{28.00}{12}
\emline{110.00}{28.00}{13}{110.00}{8.00}{14}
\emline{110.00}{8.00}{15}{70.00}{8.00}{16}
\emline{70.00}{23.00}{17}{110.00}{23.00}{18}
\emline{70.00}{18.00}{19}{110.00}{18.00}{20}
\emline{70.00}{13.00}{21}{110.00}{13.00}{22}
\emline{80.00}{28.00}{23}{80.00}{8.00}{24}
\emline{90.00}{28.00}{25}{90.00}{8.00}{26}
\end{picture}

$$ds^2=(H_1H_2)^{3\over8}(U_1U_2)^{5\over8}[-(H_1H_2U_1U_2)^{-1}dy^2+
(H_1U_2)^{-1}dy_1^2+(H_1U_1)^{-1}dy_2^2+(H_2U_2)^{-1}dy_3^2+$$
\begin{equation}
\label{D10em1}
+(H_2U_1)^{-1}dy_4^2+(U_1U_2)^{-1}(dy_5^2+dy_6^2)+
dx^{\alpha}dx^{\alpha}]
\end{equation}

\begin{equation}
{\cal
A}_8=4\pi L^6\sum_{c}(Q_{1c}Q_{2c}P_{1c}P_{2c})^{1/2}
\end{equation}

%%%%%%%%%%%%%%%%%%%%%%%%%%%%%%%%%%         END OF 9_1

%\input{pic9_2.pic}
%%%%%%%%%%%%%%%%%%%%%%%%%%%%%%%          PIC9_2.PIC
\special{em:linewidth 0.4pt}
\unitlength 1.00mm
\linethickness{0.4pt}
\begin{picture}(120.00,13.00)(0.00,5.00)
\emline{120.00}{18.00}{1}{120.00}{18.00}{2}
\emline{70.00}{8.00}{3}{70.00}{18.00}{4}
\emline{105.00}{18.00}{5}{105.00}{8.00}{6}
\emline{90.00}{18.00}{7}{90.00}{8.00}{8}
\emline{75.00}{18.00}{9}{75.00}{8.00}{10}
\put(72.67,15.33){\circle{2.11}}
\put(77.67,15.00){\circle{2.11}}
\put(82.33,15.33){\circle{2.11}}
\put(87.33,15.33){\circle{2.11}}
\put(107.67,10.67){\circle*{2.11}}
\put(87.33,10.67){\circle*{2.11}}
\put(92.67,10.67){\circle*{2.11}}
\put(98.00,10.67){\circle*{2.11}}
\put(39.67,13.00){\makebox(0,0)[cc]{$d=4,\;\alpha=\pm\sqrt{2},\;q=3$}}
\emline{70.33}{18.00}{11}{110.00}{18.00}{12}
\emline{110.00}{18.00}{13}{110.00}{8.00}{14}
\emline{110.00}{8.00}{15}{70.00}{8.00}{16}
\emline{70.00}{13.00}{17}{110.00}{13.00}{18}
\emline{85.00}{18.00}{19}{85.00}{8.00}{20}
\end{picture}

$$ds^2=(HU)^{1\over3}
[(HU)^{-{2\over3}}(-dy^2+dy_1^2+dy_2^2)+$$
\begin{equation}
+H^{-{2\over3}}dy_3^2+U^{-{2\over3}}dy_4^2+dx^{\alpha}dx^{\alpha}]
\end{equation}

In this case $q=3$, therefore the entropy is zero.

%%%%%%%%%%%%%%%%%%%%%%%%%%%%%%%%%%%%%%%%%%%%%         END OF 9_2

%\input{pic9_3.pic}
%%%%%%%%%%%%%%%%%%%%%%%%%%%%%%%%%%         PIC9_3.PIC
\special{em:linewidth 0.4pt}
\unitlength 1.00mm
\linethickness{0.4pt}
\begin{picture}(110.00,15.33)(0.00,15.00)
\emline{70.00}{8.00}{1}{70.00}{28.00}{2}
\emline{75.00}{28.00}{3}{75.00}{8.00}{4}
\emline{85.00}{28.00}{5}{85.00}{8.00}{6}
\emline{95.00}{28.00}{7}{95.00}{8.00}{8}
\emline{105.00}{28.00}{9}{105.00}{8.00}{10}
\put(72.67,25.67){\circle{2.11}}
\put(77.67,25.67){\circle{2.11}}
\put(82.33,25.67){\circle{2.11}}
\put(87.33,25.67){\circle{2.11}}
\put(92.67,25.67){\circle{2.11}}
\put(72.67,20.67){\circle{2.11}}
\put(77.67,20.67){\circle{2.11}}
\put(82.33,20.67){\circle{2.11}}
\put(97.67,20.67){\circle{2.11}}
\put(102.67,20.67){\circle{2.11}}
\put(77.67,15.67){\circle*{2.11}}
\put(82.00,15.67){\circle*{2.11}}
\put(77.67,10.67){\circle*{2.11}}
\put(82.00,10.67){\circle*{2.11}}
\put(87.33,15.67){\circle*{2.11}}
\put(92.67,10.67){\circle*{2.11}}
\put(97.67,15.67){\circle*{2.11}}
\put(102.67,10.67){\circle*{2.11}}
\put(107.67,15.67){\circle*{2.11}}
\put(107.67,10.67){\circle*{2.11}}
\put(39.67,18.00){\makebox(0,0)[cc]{d=5, $\alpha=\pm{1\over 2}$, q=1, s=1}}
\emline{70.00}{28.33}{11}{110.00}{28.33}{12}
\emline{110.00}{28.33}{13}{110.00}{8.00}{14}
\emline{110.00}{8.00}{15}{70.00}{8.00}{16}
\emline{70.00}{23.00}{17}{110.00}{23.00}{18}
\emline{70.00}{18.00}{19}{110.00}{18.00}{20}
\emline{70.00}{13.00}{21}{110.00}{13.00}{22}
\emline{90.00}{28.33}{23}{90.00}{8.00}{24}
\emline{100.00}{28.33}{25}{100.00}{8.00}{26}
\end{picture}

$$ds^2=(H_1H_2)^{5\over8}(U_1U_2)^
{3\over8}[-(H_1H_2U_1U_2)^{-1}dy^2+(H_1H_2)^{-1}(dy_1^2+dy_2^2)+
(H_1U_2)^{-1}dy_3^2+$$
\begin{equation}
\label{D10em3}
+(H_1U_1)^{-1}dy_4^2+(H_2U_2)^{-1}dy_5^2+(H_2U_1)^{-1}dy_6^2+dx^{\alpha}dx^{\alpha}]
\end{equation}

\begin{equation}
{\cal
A}_8=4\pi L^6\sum_{c}(P_{1c}P_{2c}Q_{1c}Q_{2c})^{1/2}
\end{equation}
%%%%%%%%%%%%%%%%%%%%%%%%%%%%%%          END OF 9_3

 \subsubsection{D=18, $d$-form, and $\alpha=0$}

Let us demonstrate our method in the complicated case $D=18$. For simplicity
we consider only the 'electric' ansatz.
\par
%\input{pic13_1.pic}
%%%%%%%%%%%%%%%%%%%%%%%%%%%%%%%%%          PIC13_1.PIC
\special{em:linewidth 0.4pt}
\unitlength 1.00mm
\linethickness{0.4pt}
\begin{picture}(145.00,22.00)
(0.00,10.00)
\emline{65.00}{28.00}{1}{145.00}{28.00}{2}
\emline{145.00}{28.00}{3}{145.00}{8.00}{4}
\emline{145.00}{8.00}{5}{65.00}{8.00}{6}
\emline{65.00}{8.00}{7}{65.00}{28.00}{8}
\emline{65.00}{23.00}{9}{145.00}{23.00}{10}
\emline{65.00}{18.00}{11}{145.00}{18.00}{12}
\emline{65.00}{13.00}{13}{145.00}{13.00}{14}
\emline{70.00}{28.00}{15}{70.00}{8.00}{16}
\emline{85.00}{28.00}{17}{85.00}{8.00}{18}
\emline{100.00}{28.00}{19}{100.00}{8.00}{20}
\emline{115.00}{28.00}{21}{115.00}{8.00}{22}
\emline{130.00}{28.00}{23}{130.00}{8.00}{24}
\put(67.67,25.33){\circle{2.11}}
\put(72.67,25.33){\circle{2.11}}
\put(77.67,25.33){\circle{2.11}}
\put(82.33,25.33){\circle{2.11}}
\put(67.67,20.67){\circle{2.11}}
\put(87.67,20.67){\circle{2.11}}
\put(92.67,20.67){\circle{2.11}}
\put(97.33,20.67){\circle{2.11}}
\put(67.67,15.67){\circle{2.11}}
\put(102.67,15.67){\circle{2.11}}
\put(107.67,15.67){\circle{2.11}}
\put(112.67,15.67){\circle{2.11}}
\put(67.67,10.67){\circle{2.11}}
\put(117.67,10.67){\circle{2.11}}
\put(122.67,10.67){\circle{2.11}}
\put(127.67,10.67){\circle{2.11}}
\put(35.00,18.00){\makebox(0,0)[cc]{$d=4,\;\alpha=0,\;q=1,\;s=3$}}
\end{picture}

$$ds^2=(H_1H_2H_3H_4)^{1\over6}[-(H_1H_2H_3H_4)^{-{2\over3}}dy^2+
(H_1)^{-{2\over3}}(dy_1^2+dy_2^2+dy_3^2)+$$
$$+(H_2)^{-{2\over3}}(dy_4^2+dy_5^2+dy_6^2)
+(H_3)^{-{2\over3}}(dy_7^2+dy_8^2+dy_9^2)+$$
\begin{equation}
\label{D18e1}
+(H_4)^{-{2\over3}}(dy_{10}^2+dy_{11}^2+dy_{12}^2)+dx^\alpha dx^\alpha]
\end{equation}

\begin{equation}
{\cal A}_{16}=\omega_4L^{12}\sum_{c}(Q_{1c}Q_{2c}Q_{3c}Q_{4c})^{1/3}
\end{equation}

%%%%%%%%%%%%%%%%%%%%%%%%%%%%          END OF 13_1

\par
%\input{pic13_2.pic}

%%%%%%%%%%%%%%%%%%%%%%%%%%%%%%%%%       PIC13_2.PIC
\special{em:linewidth 0.4pt}
\unitlength 1.00mm
\linethickness{0.4pt}
\begin{picture}(145.00,48.00)(0.00,10.00)
\emline{65.00}{48.00}{1}{145.00}{48.00}{2}
\emline{145.00}{48.00}{3}{145.00}{18.00}{4}
\emline{145.00}{18.00}{5}{65.00}{18.00}{6}
\emline{65.00}{18.00}{7}{65.00}{48.00}{8}
\emline{70.00}{48.00}{9}{70.00}{18.00}{10}
\emline{65.00}{43.00}{11}{145.00}{43.00}{12}
\emline{65.00}{38.00}{13}{145.00}{38.00}{14}
\emline{65.33}{33.00}{15}{145.00}{33.00}{16}
\emline{65.00}{28.00}{17}{145.00}{28.00}{18}
\emline{65.00}{23.00}{19}{145.00}{23.00}{20}
\put(67.67,45.33){\circle{2.11}}
\put(72.67,45.33){\circle{2.11}}
\put(77.67,45.33){\circle{2.11}}
\put(82.33,45.33){\circle{2.11}}
\put(87.67,45.33){\circle{2.11}}
\put(92.67,45.33){\circle{2.11}}
\put(97.33,45.33){\circle{2.11}}
\put(102.67,45.33){\circle{2.11}}
\put(67.67,40.67){\circle{2.11}}
\put(72.67,40.67){\circle{2.11}}
\put(77.67,40.67){\circle{2.11}}
\put(82.33,40.67){\circle{2.11}}
\put(107.67,40.67){\circle{2.11}}
\put(112.67,40.67){\circle{2.11}}
\put(117.67,40.67){\circle{2.11}}
\put(122.67,40.67){\circle{2.11}}
\put(67.67,35.33){\circle{2.11}}
\put(72.67,35.33){\circle{2.11}}
\put(77.67,35.33){\circle{2.11}}
\put(87.67,35.33){\circle{2.11}}
\put(107.67,35.33){\circle{2.11}}
\put(127.67,35.33){\circle{2.11}}
\put(132.67,35.33){\circle{2.11}}
\put(138.00,35.33){\circle{2.11}}
\emline{140.00}{48.00}{21}{140.00}{18.00}{22}
\put(67.67,30.67){\circle{2.11}}
\put(72.67,30.67){\circle{2.11}}
\put(92.67,30.67){\circle{2.11}}
\put(97.33,30.67){\circle{2.11}}
\put(112.67,30.67){\circle{2.11}}
\put(117.67,30.67){\circle{2.11}}
\put(127.67,30.67){\circle{2.11}}
\put(132.67,30.67){\circle{2.11}}
\put(67.67,25.33){\circle{2.11}}
\put(82.33,25.33){\circle{2.11}}
\put(92.67,25.33){\circle{2.11}}
\put(102.67,25.33){\circle{2.11}}
\put(107.67,25.33){\circle{2.11}}
\put(117.67,25.33){\circle{2.11}}
\put(127.67,25.33){\circle{2.11}}
\put(138.00,25.33){\circle{2.11}}
\put(67.67,20.33){\circle{2.11}}
\put(87.67,20.33){\circle{2.11}}
\put(97.33,20.33){\circle{2.11}}
\put(103.00,20.33){\circle{2.11}}
\put(107.67,20.33){\circle{2.11}}
\put(117.67,20.33){\circle{2.11}}
\put(122.67,20.33){\circle{2.11}}
\put(132.67,20.33){\circle{2.11}}
\emline{75.00}{48.00}{23}{75.00}{18.00}{24}
\emline{80.00}{48.00}{25}{80.00}{18.00}{26}
\emline{85.00}{48.00}{27}{85.00}{17.67}{28}
\emline{85.00}{17.67}{29}{85.00}{18.00}{30}
\emline{85.00}{18.00}{31}{85.00}{18.00}{32}
\emline{85.00}{18.00}{33}{85.00}{18.00}{34}
\emline{90.00}{48.00}{35}{90.00}{18.00}{36}
\emline{95.00}{48.00}{37}{95.00}{18.00}{38}
\emline{100.00}{48.00}{39}{100.00}{18.00}{40}
\emline{105.00}{48.00}{41}{105.00}{18.00}{42}
\emline{110.00}{48.00}{43}{110.00}{18.00}{44}
\emline{115.00}{48.00}{45}{115.00}{18.00}{46}
\emline{120.00}{48.00}{47}{120.00}{18.00}{48}
\emline{125.00}{48.00}{49}{125.00}{18.00}{50}
\emline{130.00}{48.00}{51}{130.00}{18.00}{52}
\emline{135.00}{48.00}{53}{135.00}{18.00}{54}
\put(35.00,33.00){\makebox(0,0)[cc]{d=8, $\alpha=0$, q=1, s=1}}
\emline{65.00}{18.00}{55}{65.00}{8.00}{56}
\emline{65.00}{8.00}{57}{145.00}{8.00}{58}
\emline{145.00}{8.00}{59}{145.00}{18.00}{60}
\emline{65.00}{13.00}{61}{145.00}{13.00}{62}
\emline{70.00}{18.00}{63}{70.00}{8.00}{64}
\emline{75.00}{18.00}{65}{75.00}{8.00}{66}
\emline{80.00}{18.00}{67}{80.00}{8.00}{68}
\emline{85.00}{18.00}{69}{85.00}{8.00}{70}
\emline{90.00}{18.00}{71}{90.00}{8.00}{72}
\emline{95.00}{18.00}{73}{95.00}{8.00}{74}
\emline{100.00}{18.00}{75}{100.00}{8.00}{76}
\emline{105.00}{18.00}{77}{105.00}{8.00}{78}
\emline{110.00}{18.00}{79}{110.00}{8.00}{80}
\emline{115.00}{18.00}{81}{115.00}{8.00}{82}
\emline{120.00}{18.00}{83}{120.00}{8.00}{84}
\emline{125.00}{18.00}{85}{125.00}{8.00}{86}
\emline{130.00}{17.67}{87}{130.00}{8.00}{88}
\emline{135.00}{18.00}{89}{135.00}{8.00}{90}
\emline{140.33}{18.00}{91}{140.33}{18.00}{92}
\emline{140.00}{18.00}{93}{140.00}{8.00}{94}
\put(67.67,15.33){\circle{2.11}}
\put(77.67,15.33){\circle{2.11}}
\put(92.67,15.33){\circle{2.11}}
\put(102.67,15.33){\circle{2.11}}
\put(112.67,15.33){\circle{2.11}}
\put(122.67,15.33){\circle{2.11}}
\put(132.67,15.33){\circle{2.11}}
\put(138.00,15.33){\circle{2.11}}
\put(67.67,10.67){\circle{2.11}}
\put(82.33,10.67){\circle{2.11}}
\put(87.67,10.67){\circle{2.11}}
\put(97.33,10.67){\circle{2.11}}
\put(112.67,10.67){\circle{2.11}}
\put(122.67,10.67){\circle{2.11}}
\put(127.67,10.67){\circle{2.11}}
\put(138.00,10.67){\circle{2.11}}
\end{picture}

$$ds^2=(H_1H_2H_3H_4H_5H_6H_7H_8)^{1\over
4}[-(H_1H_2H_3H_4H_5H_6H_7H_8)^{-{1\over
2}}dy^2+(H_1H_2H_3H_4)^{-{1\over 2}}dy_1^2+$$
$$+(H_1H_2H_3H_7)^{-{1\over 2}}dy_2^2+
+(H_1H_2H_5H_8)^{-{1\over
2}}dy_3^2+(H_1H_3H_6H_8)^{-{1\over2}}dy_4^2+$$
 $$+(H_1H_4H_5H_7)^{-{1\over2}}dy_5^2+
+(H_1H_4H_6H_8)^{-{1\over2}}dy_6^2+
(H_1H_5H_6H_7)^{-{1\over2}}dy_7^2+(H_2H_3H_5H_6)^{-{1\over2}}dy_8^2+$$
$$+(H_2H_4H_7H_8)^{-{1\over2}}dy_9^2+(H_2H_4H_5H_6)^{-{1\over2}}dy_{10}^2+
(H_2H_6H_7H_8)^{-{1\over2}}dy_{11}^2+
(H_3H_4H_5H_8)^{-{1\over2}}dy_{12}^2+$$
\begin{equation}
\label{D18e2}
(H_3H_4H_6H_7)^{-{1\over2}}dy_{13}^2+
(H_3H_5H_7H_8)^{-{1\over2}}dy_{14}^2+dx^\alpha dx^\alpha]
\end{equation}

\begin{equation}
{\cal
A}_{16}=4\pi L^{14}\sum_{c}(Q_{1c}Q_{2c}Q_{3c}Q_{4c}Q_{5c}Q_{6c}
Q_{7c}Q_{8c})^{1/4}
\end{equation}

%%%%%%%%%%%%%%%%%%%%%%%%%%%%%                   END OF 13_2

\par
%\input{pic13_3.pic}
%%%%%%%%%%%%%%%%%%%%%%%%%%%%%%%%          PIC13_3.PIC
\special{em:linewidth 0.4pt}
\unitlength 1.00mm
\linethickness{0.4pt}
\begin{picture}(145.00,20.00)(0.00,12.00)
\emline{65.00}{28.00}{1}{145.00}{28.00}{2}
\emline{145.00}{28.00}{3}{145.00}{8.00}{4}
\emline{145.00}{8.00}{5}{65.00}{8.00}{6}
\emline{65.00}{8.00}{7}{65.00}{28.00}{8}
\emline{65.00}{23.00}{9}{145.00}{23.00}{10}
\emline{65.00}{18.00}{11}{145.00}{18.00}{12}
\emline{65.00}{13.00}{13}{145.00}{13.00}{14}
\emline{80.00}{28.00}{15}{80.00}{8.00}{16}
\emline{95.00}{28.00}{17}{95.00}{8.00}{18}
\emline{110.00}{28.00}{19}{110.00}{8.00}{20}
\emline{140.00}{28.00}{21}{140.00}{8.00}{22}
\emline{125.00}{28.00}{23}{125.00}{8.00}{24}
\put(67.67,25.33){\circle{2.11}}
\put(72.67,25.33){\circle{2.11}}
\put(77.67,25.33){\circle{2.11}}
\put(82.33,25.33){\circle{2.11}}
\put(87.67,25.33){\circle{2.11}}
\put(92.67,25.33){\circle{2.11}}
\put(97.33,25.33){\circle{2.11}}
\put(102.67,25.33){\circle{2.11}}
\put(107.67,25.33){\circle{2.11}}
\put(112.67,25.33){\circle{2.11}}
\put(117.67,25.33){\circle{2.11}}
\put(122.67,25.33){\circle{2.11}}
\put(67.67,20.33){\circle{2.11}}
\put(72.67,20.33){\circle{2.11}}
\put(77.67,20.33){\circle{2.11}}
\put(82.33,20.33){\circle{2.11}}
\put(87.67,20.33){\circle{2.11}}
\put(92.67,20.33){\circle{2.11}}
\put(97.33,20.33){\circle{2.11}}
\put(102.67,20.33){\circle{1.89}}
\put(107.67,20.33){\circle{2.00}}
\put(67.67,15.33){\circle{2.11}}
\put(72.67,15.33){\circle{2.11}}
\put(77.67,15.33){\circle{2.11}}
\put(82.33,15.33){\circle{2.11}}
\put(87.67,15.33){\circle{2.11}}
\put(92.67,15.33){\circle{2.11}}
\put(97.33,10.33){\circle{2.11}}
\put(102.67,10.33){\circle{2.11}}
\put(107.67,10.33){\circle{2.11}}
\put(112.67,10.33){\circle{2.11}}
\put(117.67,10.33){\circle{2.11}}
\put(122.67,10.33){\circle{2.11}}
\put(127.67,20.33){\circle{2.11}}
\put(132.67,20.33){\circle{2.11}}
\put(138.00,20.33){\circle{2.11}}
\put(67.67,10.33){\circle{2.11}}
\put(72.67,10.33){\circle{2.11}}
\put(77.67,10.33){\circle{2.11}}
\put(127.67,10.33){\circle{2.11}}
\put(132.67,10.33){\circle{2.11}}
\put(138.00,10.33){\circle{2.11}}
\put(112.67,15.33){\circle{2.11}}
\put(117.67,15.33){\circle{2.11}}
\put(122.67,15.33){\circle{2.11}}
\put(127.33,15.33){\circle{2.11}}
\put(132.67,15.33){\circle{2.11}}
\put(137.67,15.33){\circle{2.11}}
\put(35.00,18.00){\makebox(0,0)[cc]{d=12, $\alpha=0$, q=3, s=1}}
\end{picture}

$$ds^2=(H_1H_2H_3H_4)^{1\over 2}[(H_1H_2H_3H_4)^{-{2\over 3}}
(-dy^2+dy_1^2+dy_2^2)+(H_1H_2H_3)^{-{2\over 3}}(dy_3^2+dy_4^2+dy_5^2)+$$
$$+(H_1H_2H_4)^{-{2\over 3}}(dy_6^2+dy_7^2+dy_8^2)+
(H_1H_3H_4)^{-{2\over 3}}(dy_9^2+dy_{10}^2+dy_{11}^2)+$$
\begin{equation}
\label{D10e3a}
+(H_2H_3H_4)^{-{2\over 3}}(dy_{12}^2+dy_{13}^2+dy_{14}^2)
+dx^\alpha dx^\alpha]
\end{equation}

In this solution $q=3$, the area of horizon is zero.
%%%%%%%%%%%%%%%%%%%%%%%%%%                 END OF 13_3

\newpage
\section { Gravity coupled with $d_1$-form, $d_2$-form
and dilaton}
\setcounter{equation}{0}
In this section we will consider soliton solution of the form (\ref{3})
for the theory
\begin{equation}
 I=\frac{1}{2\kappa ^{2}}\int d^{D}x\sqrt{-g}
 (R- \frac{1}{2}(\nabla \phi)^{2}-\frac{e^{-\alpha \phi}}
{2(d_{1}+1)!}F^{2}_{d_{1}+1}- \frac{e^{-\beta
\phi}}{2(d_{2}+1)!}F^{2}_{d_{2}+1}),
                                           \label{2}
 \end{equation}

  We will follow  the same strategy of finding solutions
 as before. For simplicity let us present the main steps of calculations only
 for pure
'electric' field but write only the answer for a general
'electric' + 'magnetic' ansatz.

For $d_1$ and  $d_2$-form we are going to consider a class of ansatzes
corresponding to $E_1$ and $E_{2}$ electric charges. Each ansatz we will
connect
with two incidence matrices  $\Delta_{ai}^{(1)},\;a=1..E_1,\;i=1..N$,
$\Delta_{bi}^{(2)},\;b=1..E_2,\;i=1..N$, which apart the properties 1), 2),
3) and 5)
mentioned in the previous section satisfy the following conditions

4)$\;\sum_{i=1}^N r_i\Delta_{ai}^{(1)}=d_1-q\,$ for all $a$;

$\;\sum_{i=1}^N r_i\Delta_{ai}^{(2)}=d_2-q\,$ for all $a$;

Electric fields are assumed in the form
  \begin{equation}
 \label{a1}
{\cal
  A}^{(1)}=\sum_{a=1}^{E_1}h_a^{(1)}e^{C_a(x)^{(1)}}
\omega ^{q}_{0}\prod _{i|\Delta ^{(1)}_{ai}=1}\wedge~\omega _{i}
\end{equation}

  \begin{equation}
 \label{a2}
{\cal A}^{(2)}=\sum_{a=1}^{E_2}h_a^{(2)}e^{C_a(x)^{(2)}}
\omega ^{q}_{0}\prod _{i|\Delta ^{(2)}_{ai}=1}\wedge~\omega _{i}
\end{equation}

Components of the energy-momentum tensor for $d_1$-form and $d_2$-form are
very similar except that in the exponent
corresponding to the $d_{2}$-field new parameter
$\beta$ enters instead of $\alpha $ .

"No-force" conditions in this case have the form
 \begin{equation}
 \label{nfe1}
-\alpha\phi-2qA-2\sum_{i=1}^N\Delta_{ai}^{(1)}r_iF_i+2C_a^{(1)}=0,
\;i=1..E_1
\end{equation}

 \begin{equation}
 \label{nfe2}
-\beta\phi-2qA-2\sum_{i=1}^N\Delta_{ai}^{(2)}r_iF_i+2C_a^{(2)}=0,
\;i=1..E_2
\end{equation}

 From Maxwell's equations under conditions
(\ref{nfe1}) and (\ref{nfe2}) we get

$$\Delta C_a^{(1)}=(\partial C_a^{(1)})^2, ~~~
\Delta C_a^{(2)}=(\partial
C_a^{(2)})^2$$

Having in our disposal these expressions the
equation of motion for the dilaton as well as the Einstein equations
(apart from the non-diagonal part for $\gamma$-components  can
be solved if one assume that

 \begin{equation}
 \label{ea}
A=\sum_{a=1}^{E_1}t^{(1)}{h_a^{(1)}}^2C_a^{(1)}
+\sum_{a=1}^{E_2}t^{(2)}{h_a^{(2)}}^2C_a^{(2)}
\end{equation}

 \begin{equation}
 \label{eb}
F_i=\sum_{a=1}^{E_1}{h_a^{(1)}}^2C_a^{(1)}({1\over
2}\Delta_{ai}^{(1)}-u^{(1)})+\sum_{a=1}^{E_2}{h_a^{(2)}}^2C_a^{(2)}({1\over
2}\Delta_{ai}^{(2)}-u^{(2)})
\end{equation}

 \begin{equation}
 \label{ef}
B=-u^{(1)}\sum_{a=1}^{E_1}{h_a^{(1)}}^2C_a^{(1)}
-u^{(2)}\sum_{a=1}^{E_2}{h_a^{(2)}}^2C_a^{(2)}
\end{equation}

 \begin{equation}
 \label{ep}
\phi={\alpha\over 2}\sum_{a=1}^{E_1}{h_a^{(1)}}^2
C_a^{(1)}+{\beta\over 2}\sum_{a=1}^{E_2}{h_a^{(2)}}^2
C_a^{(2)}
\end{equation}
where $t^{(1)},\;t^{(2)},\;u^{(1)}$ and $u^{(2)}$ are given by
\begin{equation}
t^{(1)}={D-2-d_1\over 2(D-2)}
\end{equation}
\begin{equation}
t^{(2)}={D-2-d_2\over 2(D-2)}
\end{equation}
\begin{equation}
u^{(1)}={d_1\over 2(D-2)}
\end{equation}
\begin{equation}
u^{(2)}={d_2\over 2(D-2)}
\end{equation}

Having the formulae (\ref{ea}), (\ref{eb}) and (\ref{ef})
one can immediate verify that the relation (\ref{1.3})
holds.

In the same way as it has been done in Section 2
one can find that consistency conditions of equations
(\ref{ea}), (\ref{eb}), \ref{ef}) and (\ref{ep}) with (\ref{nf1}) and
(\ref{nf2})
force us to put the following restrictions

 \begin{equation}
 \label{3.1}
(D-2)({\alpha^2\over
2}+q)-d_1^2+(D-2)\sum_{i=1}^Nr_i\Delta_{ai}^{(1)}\Delta_{a'i}^{(1)}=0,
~~a\not=a'
\end{equation}

\begin{equation}
 \label{3.2}
({\alpha\beta\over2}+q)(D-2)-d_1d_2+(D-2)
\sum_{i=1}^Nr_i\Delta_{ai}^{(1)}\Delta_{a'i}^{(2)}=0
\end{equation}

\begin{equation}
 \label{3.3}
(D-2)({\beta^2\over
2}+q)-d_2^2+(D-2)\sum_{i=1}^Nr_i\Delta_{ai}^{(2)}\Delta_{a'i}^{(2)}=0
~~~a\not=a'
\end{equation}

and take $h_{a}$ in the form

\begin{equation}
 \label{3.4}
{h_a^{(1)}}^2=\sigma^{(1)}={1\over t^{(1)}d_1+{\alpha^2\over4}} ,~~~
{h_a^{(2)}}^2=\sigma^{(2)}={1\over t^{(2)}d_2+{\beta^2\over
4}}
\end{equation}

One can check that if these conditions are satisfied then the rest of the Einstein
equations are hold. Introducing the notations $\alpha=\alpha^{(1)},\;
\beta=\alpha^{(2)}$ we can rewrite (\ref{3.1}) --(\ref{3.3}) as
\begin{equation}
\label{condee}
(1-\delta_{IJ}\delta_{aa'})\{ {\alpha^{(I)}\alpha^{(J)}\over
2}+q-{d_Id_J\over D-2}+\sum_{i=1}^Nr_i\Delta_{ai}^{(I)}\Delta_{a'i}^{(J)}
\}=0,
\;I,\;J=1,\;2
\end{equation}

Let us present the  metric corresponding to
 this solution

$$ds^{2}=
(H_{1}^{(1)}H_{2}^{(1)}...H_{E_1}^{(1)})^{2u^{(1)}\sigma^{(1)}}
(H_{1}^{(2)}H_{2}^{(2)}...H_{E_2}^{(2)})^{2u^{(2)}\sigma^{(2)}}$$
$$\{(H_{1}^{(1)}H_{2}^{(1)}...H_{E_1}^{(1)})^{-\sigma^{(1)}}
(H_{1}^{(2)}H_{2}^{(2)}...H_{E_2}^{(2)})^{-\sigma^{(2)}}
\eta_{\mu \nu} dy^{\mu}dy^{\nu}+$$
\begin{equation}
+\sum_{i}(\prod _{a}{H_{a}^{(1)}}^{-\Delta
_{ai}^{(1)}})^{\sigma^{(1)}} (\prod
_{a}{H_{a}^{(2)}}^{-\Delta_{ai}^{(2)}})^{\sigma^{(2)}} \sum
_{m_{i}=1}^{r_{i}}dz^{m_{i}}dz^{m_{i}} +\sum _{\alpha
}dx^{\alpha}dx^{\alpha}],
                                    \label{8''}
\end{equation}
In the most general case with electric and magnetic fields we have additional
restrictions on the incidence matrices

\begin{equation}
\label{condmm}
(1-\delta_{IJ}\delta_{bb'})\{ {\alpha^{(I)}\alpha^{(J)}\over
2}+s-{d_Id_J\over D-2}+\sum_{i=1}^Nr_i\Lambda_{bi}^{(I)}\Lambda_{b'i}^{(J)}
\}=0,
\end{equation}

\begin{equation}
\label{condem}
{\alpha^{(I)}\alpha^{(J)}\over
2}+s-{d_Id_J\over D-2}+\sum_{i=1}^Nr_i\Delta_{ai}^{(I)}\Lambda_{b'i}^{(J)}
=0,
\;I,\;J=1,\;2
\end{equation}

and
\begin{equation}
{v_m^{(I)}}^2={v^{(I)}}^2={1\over
t^{(I)}d_I +{\alpha_I^2\over 4}},\;I=1,\;2
\end{equation}

The metric has the form
$$ds^{2}=
(H_{1}^{(1)}H_{2}^{(1)}...H_{E_1}^{(1)})^{2u^{(1)}\sigma^{(1)}}
(U_{1}^{(1)}U_{2}^{(1)}...U_{M_1}^{(1)})^{2t^{(1)}\sigma^{(1)}}
(H_{1}^{(2)}H_{2}^{(2)}...H_{E_2}^{(2)})^{2u^{(2)}\sigma^{(2)}}$$
$$ (U_{1}^{(2)}U_{2}^{(2)}...U_{M_2}^{(2)})^{2t^{(2)}\sigma^{(2)}}
\{(H_{1}^{(1)}H_{2}^{(1)}...H_{E_1}^{(1)}
U_{1}^{(1)}U_{2}^{(1)}...U_{M_1}^{(1)})^{-\sigma^{(1)}}
(H_{1}^{(2)}H_{2}^{(2)}...H_{E_2}^{(2)}$$ $$U_{1}^{(2)}U_{2}^
{(2)}...U_{M_2}^{(2)})^{-\sigma^{(2)}}
\eta_{\mu \nu} dy^{\mu}dy^{\nu}+
\sum_{i}(\prod _{a}{H_{a}^{(1)}}^{-\Delta
_{ai}^{(1)}}\prod _{b}{U_{b}^{(1)}}^{\Lambda
_{bi}^{(1)}})^{\sigma^{(1)}}$$
\begin{equation}
\label{genmetr}
  (\prod
_{a}{H_{a}^{(2)}}^{-\Delta_{ai}^{(2)}}\prod
_{b}{U_{b}^{(2)}}^{\Lambda_{ai}^{(2)}})^{\sigma^{(2)}} \sum
_{m_{i}=1}^{r_{i}}dz^{m_{i}}dz^{m_{i}} +\sum _{\alpha
}dx^{\alpha}dx^{\alpha}],
                                    \label{}
\end{equation}

Note that in the case of one dilaton field and an arbitrary number of
$d_I$-forms, $I=1,..,S$, requirements on $2S$ incidence matrices are also
given by ~(\ref{condee}), (\ref{condmm}) and (\ref{condem})  with
$I,\;J=1,..,S$ and corresponding metric is an obvious generalization
of ~(\ref{genmetr}). Furthermore, the same method could be applied to
the case with more then one dilaton fields.

\subsection{Examples}
\subsubsection{D=11}
We present below only solutions with at list two branches fot both fields.
There are more solutions with one branch for earch field.

1) {\it ''electric'' ansatzes.}

 In this case we
get
$$ q-{d_1d_2\over
 9}+\sum_{i=1}^Nr_i\Delta_{ai}^{1}\Delta_{a'i}^{2}=0$$
and we have two types of solution
$$~$$

%\input{pic4.pic}
%%%%%%%%%%%%%%%%%%%%%%%%%%%%%%%%%%%%%%%%%%%         PIC4.PIC
\unitlength 1.00mm
\linethickness{0.4pt}
\begin{picture}(110.00,20.00)(0.00,75.00)
\put(65.00,80.00){\line(0,1){15.00}}
\put(65.00,95.00){\line(1,0){5.33}}
\put(70.33,95.00){\line(-1,0){5.33}}
\put(65.00,95.00){\line(1,0){5.33}}
\put(70.33,95.00){\line(-1,0){5.33}}
\put(67.67,92.33){\circle{2.11}}
\put(67.67,87.67){\circle{2.11}}
\put(77.33,92.33){\circle{2.11}}
\put(82.67,87.67){\circle{2.11}}
\put(87.00,87.67){\circle{2.11}}
\put(72.67,92.33){\circle{2.11}}
\put(92.33,77.67){\circle{2.11}}
\put(67.67,72.67){\circle{2.11}}
\put(82.33,77.67){\circle{2.11}}
\put(78.00,72.67){\circle{2.11}}
\put(97.67,72.67){\circle{2.11}}
\put(102.33,72.67){\circle{2.11}}
\put(67.67,77.67){\circle{2.11}}
\put(72.33,77.67){\circle{2.11}}
\put(29.00,74.67){\makebox(0,0)[cc]{$d_2=6$, q=2, s=1}}
\put(29.00,90.00){\makebox(0,0)[cc]{$d_{1}=3$, q=1, s=1}}
\put(65.00,80.00){\line(0,1){5.00}}
\put(90.00,85.00){\line(0,-1){0.33}}
\put(65.00,80.00){\line(0,-1){10.00}}
\put(70.00,80.00){\line(0,-1){10.00}}
\put(80.00,80.00){\line(0,-1){10.00}}
\put(90.00,79.67){\line(0,-1){9.67}}
\put(105.00,80.00){\line(0,-1){10.00}}
\put(105.00,95.00){\line(0,-1){10.00}}
\put(87.00,72.67){\circle{2.11}}
\put(92.33,72.67){\circle{2.11}}
\put(97.67,77.67){\circle{2.11}}
\put(102.33,77.67){\circle{2.11}}
\put(70.00,95.00){\line(1,0){40.00}}
\put(110.00,95.00){\line(0,-1){25.00}}
\put(110.00,70.00){\line(-1,0){45.00}}
\put(65.00,90.00){\line(1,0){45.00}}
\put(65.00,75.00){\line(1,0){45.00}}
\put(75.00,95.00){\line(0,-1){10.00}}
\put(75.00,80.00){\line(0,-1){10.00}}
\put(85.00,95.00){\line(0,-1){10.00}}
\put(85.00,80.00){\line(0,-1){10.00}}
\put(65.00,85.00){\line(1,0){45.00}}
\put(65.00,80.00){\line(1,0){45.00}}
\put(70.00,95.00){\line(0,-1){10.00}}
\put(90.00,95.00){\line(0,-1){10.00}}
\end{picture}

$$~$$

$$ds^2=(H_1^{(1)}H_2^{(1)})^{1\over3}(H_1^{(2)}H_2^{(2)})^{2\over3}
[-(H_1^{(1)}H_2^{(1)}H_1^{(2)}H_2^{(2)})^{-1}dy^2+$$
$$(H_1^{(1)}H_1^{(2)})^
{-1}dy_1^2+(H_1^{(1)}H_2^{(2)})^{-1}dy_2^2
+(H_2^{(1)}H_1^{(2)})^{-1}dy_3^2+(H_2^{(1)}H_2^{(2)})^{-1}dy_4^2+$$
\begin{equation}
\label{D11f2e}
(H_1^{(2)}H_2^{(2)})^{-1}(dy_5^2+dy_6^2+dy_7^2)+dx^{\alpha}dx^{\alpha}]
\end{equation}

\begin{equation}
{\cal A}_9=4\pi L^7
\sum_{c}(Q^{(1)}_{1c}Q^{(1)}_{2c}Q^{(2)}_{1c}Q^{(2)}_{2c})^{1/2}
\end{equation}
%%%%%%%%%%%%%%%%%%%%%%%%%%%%%                END OF 4

2) {\it ''magnetic''   ansatzes}

 In this case we get
$$s-{d_1d_2\over 9}+\sum_{i=1}^Nr_i\Lambda_{bi}^1\Lambda_{b'i}^2=0 $$
and corresponding solutions are
$$~$$

%\input{pic5.pic}
%%%%%%%%%%%%%%%%%%%%%%%%%%%%%%%             PIC5.PIC
\unitlength 1.00mm
\linethickness{0.4pt}
\begin{picture}(110.00,25.00)(0.00,95.00)
\put(29.67,114.34){\makebox(0,0)[cc]{$d_1=3$, q=1, s=1}}
\put(29.67,99.34){\makebox(0,0)[cc]{$d_2=6$, q=1, s=1}}
\put(65.00,110.00){\line(0,1){10.00}}
\put(65.00,105.00){\line(0,-1){10.00}}
\put(72.67,112.67){\circle*{2.11}}
\put(77.33,117.67){\circle*{2.10}}
\put(83.00,112.67){\circle*{2.11}}
\put(87.33,117.67){\circle*{2.11}}
\put(108.00,117.67){\circle*{2.11}}
\put(108.00,112.67){\circle*{2.11}}
\put(72.67,97.67){\circle*{2.11}}
\put(77.33,97.67){\circle*{2.11}}
\put(83.00,102.67){\circle*{2.11}}
\put(87.33,102.67){\circle*{2.11}}
\put(93.00,102.67){\circle*{2.11}}
\put(97.67,102.67){\circle*{2.11}}
\put(102.67,102.67){\circle*{2.11}}
\put(108.00,102.67){\circle*{2.11}}
\put(108.00,97.67){\circle*{2.11}}
\put(93.00,98.00){\circle*{2.11}}
\put(97.67,98.00){\circle*{2.11}}
\put(102.67,98.00){\circle*{2.11}}
\put(65.00,120.00){\line(1,0){45.00}}
\put(110.00,120.00){\line(0,-1){25.00}}
\put(110.00,95.00){\line(-1,0){45.00}}
\put(65.00,115.00){\line(1,0){45.00}}
\put(65.00,100.00){\line(1,0){45.00}}
\put(75.00,120.00){\line(0,-1){10.00}}
\put(75.00,105.00){\line(0,-1){10.00}}
\put(85.00,119.67){\line(0,-1){9.67}}
\put(85.00,105.00){\line(0,-1){10.00}}
\put(65.00,110.00){\line(1,0){45.00}}
\put(65.00,105.00){\line(1,0){45.00}}
\put(65.00,110.00){\line(0,-1){5.00}}
\put(80.00,120.00){\line(0,-1){10.00}}
\put(105.00,120.00){\line(0,-1){10.00}}
\put(105.00,105.00){\line(0,-1){10.00}}
\put(80.00,105.00){\line(0,-1){10.00}}
\put(90.00,120.00){\line(0,-1){10.00}}
\put(90.00,105.00){\line(0,-1){10.00}}
\put(70.00,120.00){\line(0,-1){10.00}}
\put(70.00,105.00){\line(0,-1){10.00}}
\end{picture}

$$ds^2=(U_1^{(1)}U_2^{(1)})^{2\over3}(U_1^{(2)}U_2^{(2)})^{1\over3}
[-U_1^{(1)}U_2^{(1)}U_1^{(2)}U_2^{(2)}dy^2+$$
$$+(U_1^{(1)}U_1^{(2)})^{-1} dy_1^2+(U_2^{(1)}U_1^{(2)})^{-1}dy_2^2+$$
\begin{equation}
\label{D11f2m}
+(U_1^{(1)}U_2^{(2)})^{-1}dy_3^2+(U_2^{(1)}U_2^{(2)})^{-1}dy_4^2+
(U_1^{(1)}U_2^{(1)})^{-1}(dy_5^2+dy_6^2+dy_7^2)+dx^{\alpha}dx^{\alpha}]
\end{equation}

\begin{equation}
{\cal A}_9=4\pi L^7\sum_{c}(P^{(1)}_{1c}P^{(1)}_{2c}P^{(2)}_{1c}
P^{(2)}_{2c})^{1/2}
\end{equation}
%%%%%%%%%%%%%%%%%%%%%%%%%%%%%        END OF 5

 3){\it ''electric'' and ''magnetic''   ansatzes.}

New conditions are
$$-{d_1d_2\over 9}+\sum_{i=1}^Nr_i\Lambda_{bi}^1\Delta_{ai}^2=0 $$
$$-{d_1d_2\over 9}+\sum_{i=1}^Nr_i\Lambda_{bi}^2\Delta_{ai}^1=0 $$
We obtained three classes of solutions
$$~$$
%\input{pic6_1.pic}
%%%%%%%%%%%%%%%%%%%%%%%%%%%%%%%%%           PIC6_1.PIC
\unitlength 1.00mm
\linethickness{0.4pt}
\begin{picture}(110.00,20.00)(0.00,12.00)
\put(65.00,8.00){\line(0,1){0.00}}
\put(65.00,8.00){\line(0,1){25.00}}
\put(70.00,32.67){\line(0,1){0.33}}
\put(70.00,33.00){\line(0,1){0.00}}
\put(70.00,33.00){\line(0,1){0.00}}
\put(70.00,33.00){\line(0,1){0.00}}
\put(70.00,33.00){\line(0,1){0.00}}
\put(70.00,33.00){\line(0,1){0.00}}
\put(70.00,33.00){\line(0,1){0.00}}
\put(70.00,33.00){\line(0,-1){15.00}}
\put(70.00,12.67){\line(0,-1){4.67}}
\put(80.00,33.00){\line(0,-1){14.67}}
\put(80.00,13.00){\line(0,-1){5.00}}
\put(90.00,33.00){\line(0,-1){15.00}}
\put(90.00,13.00){\line(0,-1){5.00}}
\put(105.00,33.00){\line(0,-1){15.00}}
\put(105.00,13.00){\line(0,-1){5.00}}
\put(67.67,30.67){\circle{2.11}}
\put(73.00,30.67){\circle{2.11}}
\put(77.33,30.67){\circle{2.11}}
\put(67.67,25.67){\circle{2.11}}
\put(82.67,25.67){\circle{2.11}}
\put(87.33,25.67){\circle{2.11}}
\put(77.33,20.67){\circle*{2.11}}
\put(87.33,20.67){\circle*{2.11}}
\put(107.67,20.67){\circle*{2.11}}
\put(67.67,10.67){\circle{2.11}}
\put(77.33,10.67){\circle{2.11}}
\put(87.33,10.67){\circle{2.11}}
\put(92.67,10.67){\circle{2.11}}
\put(97.67,10.67){\circle{2.11}}
\put(102.67,10.67){\circle{2.11}}
\put(29.67,25.67){\makebox(0,0)[cc]{$d_1=3$, q=1, s=1}}
\put(29.33,10.67){\makebox(0,0)[cc]{$d_2=6$, q=1,s=1}}
\put(65.00,33.00){\line(1,0){45.00}}
\put(110.00,33.00){\line(0,-1){25.00}}
\put(110.00,8.00){\line(-1,0){45.00}}
\put(65.00,28.00){\line(1,0){45.00}}
\put(65.00,23.00){\line(1,0){45.00}}
\put(75.00,33.00){\line(0,-1){15.00}}
\put(75.00,13.00){\line(0,-1){5.00}}
\put(85.00,33.00){\line(0,-1){15.00}}
\put(85.00,13.00){\line(0,-1){5.00}}
\put(65.00,17.67){\line(1,0){45.00}}
\put(65.00,13.00){\line(1,0){45.00}}
\end{picture}

$$ds^2=(H_1^{(1)}H_2^{(1)})^{1\over3}(UH^{(2)})^{2\over3}
[-(H_1^{(1)}H_2^{(1)}UH^{(2)})^{-1}dy^2+(H_1^{(1)}U)^{-1}dy_1^2+
(H_1^{(1)}H^{(2)})^{-1}dy_2^2+$$
\begin{equation}
\label{D11f2em1}
+(H_2^{(1)}U)^{-1}dy_3^2+(H_2^{(1)}H^{(2)})^{-1}dy_4^2+
(UH^{(2)})^{-1}(dy_5^2+dy_6^2+dy_7^2)+dx^{\alpha}dx^{\alpha}]
\end{equation}

%\begin{equation}
%{\cal A}=4\pi\sum_{c_0}(Q^{(1)}_{1c_0}Q^{(1)}_{2c_0}Q^{(2)}_{c_0}
%P_{c_0})^{1/2}
%\end{equation}

%%%%%%%%%%%%%%%%%%%%%%%%%%%                END OF 6_1

%\input{pic6_2.pic}
%%%%%%%%%%%%%%%%%%%%%%%%%%%%%%%          PIC6_2.PIC
\unitlength 1.00mm
\linethickness{0.4pt}
\begin{picture}(110.00,33.00)(0.00,5.00)
\put(65.00,8.00){\line(0,1){0.00}}
\put(65.00,8.00){\line(0,1){25.00}}
\put(70.00,32.67){\line(0,1){0.33}}
\put(70.00,33.00){\line(0,1){0.00}}
\put(70.00,33.00){\line(0,1){0.00}}
\put(70.00,33.00){\line(0,1){0.00}}
\put(70.00,33.00){\line(0,1){0.00}}
\put(70.00,33.00){\line(0,1){0.00}}
\put(70.00,33.00){\line(0,1){0.00}}
\put(70.00,33.00){\line(0,-1){15.00}}
\put(70.00,12.67){\line(0,-1){4.67}}
\put(90.00,33.00){\line(0,-1){15.00}}
\put(90.00,13.00){\line(0,-1){5.00}}
\put(105.00,33.00){\line(0,-1){15.00}}
\put(105.00,13.00){\line(0,-1){5.00}}
\put(85.00,32.67){\line(0,1){0.00}}
\put(85.00,32.67){\line(0,1){0.33}}
\put(85.00,33.00){\line(0,1){0.00}}
\put(85.00,33.00){\line(0,1){0.00}}
\put(85.00,33.00){\line(0,1){0.00}}
\put(85.00,33.00){\line(0,-1){15.00}}
\put(85.00,13.00){\line(0,-1){5.00}}
\put(67.67,30.33){\circle{2.11}}
\put(73.00,30.33){\circle{2.11}}
\put(77.33,30.33){\circle{2.11}}
\put(82.00,30.33){\circle{2.11}}
\put(87.33,30.33){\circle{2.11}}
\put(92.67,30.33){\circle{2.11}}
\put(97.67,25.67){\circle{2.11}}
\put(102.67,25.67){\circle{2.11}}
\put(107.67,20.67){\circle*{2.11}}
\put(102.67,20.67){\circle*{2.11}}
\put(92.67,20.67){\circle*{2.11}}
\put(73.00,20.67){\circle*{2.11}}
\put(77.33,20.67){\circle*{2.11}}
\put(82.00,20.67){\circle*{2.11}}
\put(67.67,10.33){\circle{2.11}}
\put(92.67,10.33){\circle{2.11}}
\put(102.67,10.33){\circle{2.11}}
\put(70.00,27.67){\line(0,1){0.33}}
\put(70.00,28.00){\line(0,1){0.00}}
\put(70.00,28.00){\line(0,1){0.00}}
\put(70.00,28.00){\line(0,1){0.00}}
\put(70.00,28.00){\line(0,1){0.00}}
\put(70.00,28.00){\line(0,1){0.00}}
\put(70.00,28.00){\line(0,1){0.00}}
\put(85.00,27.67){\line(0,1){0.00}}
\put(85.00,27.67){\line(0,1){0.33}}
\put(85.00,28.00){\line(0,1){0.00}}
\put(85.00,28.00){\line(0,1){0.00}}
\put(85.00,28.00){\line(0,1){0.00}}
\put(67.67,25.33){\circle{2.11}}
\put(73.00,25.33){\circle{2.11}}
\put(77.33,25.33){\circle{2.11}}
\put(82.00,25.33){\circle{2.11}}
\put(29.33,25.33){\makebox(0,0)[cc]{$d_1=6$, q=1, s=1}}
\put(29.33,10.33){\makebox(0,0)[cc]{$d_2=3$, q=1, s=1}}
\put(65.00,33.00){\line(0,1){0.00}}
\put(65.00,33.00){\line(1,0){45.00}}
\put(110.00,33.00){\line(0,-1){25.00}}
\put(110.00,8.00){\line(-1,0){45.00}}
\put(65.00,28.00){\line(1,0){45.00}}
\put(65.00,23.00){\line(1,0){45.00}}
\put(95.00,33.00){\line(0,-1){15.00}}
\put(95.00,13.00){\line(0,-1){5.00}}
\put(100.00,33.00){\line(0,-1){15.00}}
\put(100.00,13.00){\line(0,-1){5.00}}
\put(65.00,17.67){\line(1,0){45.00}}
\put(65.00,13.00){\line(1,0){45.00}}
\end{picture}

$$ds^2=(H_1^{(1)}H_2^{(1)})^{2\over3}(UH^{(2)})^{1\over3}[-
(H_1^{(1)}H_2^{(1)}UH^{(2)})^{-1}dy^2+$$
$$+(H_1^{(1)}H_2^{(1)})^{-1}(dy_1^2+
dy_2^2+dy_3^2)+(H_1^{(1)}U)^{-1}dy_4^2+$$
\begin{equation}
+(H_1^{(1)}H^{(2)})^{-1}dy_5^2+(H_2^{(1)}U)^{-1}dy_6^2+
(H_2^{(1)}H^{(2)})^{-1}dy_7^2+dx^{\alpha}dx^{\alpha}]
\end{equation}

%\begin{equation}
%{\cal A}=4\pi\sum_{c_0}(Q^{(1)}_{1c_0}Q^{(1)}_{2c_0}Q^{(2)}_{c_0}
%P_{c_0})^{1/2}
%\end{equation}
%%%%%%%%%%%%%%%%%%%%%%%%%%%%%             END OF 6_2

%\input{pic6_3.pic}
%%%%%%%%%%%%%%%%%%%%%%%%%%%%%%%%             PIC6_3.PIC
\unitlength 1.00mm
\linethickness{0.4pt}
\begin{picture}(110.00,33.00)(0.00,5.00)
\put(65.00,8.00){\line(0,1){25.00}}
\put(70.00,33.00){\line(0,-1){10.00}}
\put(70.00,18.00){\line(0,-1){10.00}}
\put(80.00,33.00){\line(0,-1){10.00}}
\put(80.00,18.00){\line(0,-1){10.00}}
\put(90.00,33.00){\line(0,-1){10.00}}
\put(90.00,18.00){\line(0,-1){10.00}}
\put(105.00,33.00){\line(0,-1){10.00}}
\put(105.00,18.00){\line(0,-1){10.00}}
\put(67.67,30.33){\circle{2.11}}
\put(73.00,30.33){\circle{2.11}}
\put(77.33,30.33){\circle{2.11}}
\put(77.33,25.67){\circle*{2.11}}
\put(87.33,25.67){\circle*{2.11}}
\put(107.67,25.67){\circle*{2.11}}
\put(67.67,15.33){\circle{2.11}}
\put(77.33,15.33){\circle{2.11}}
\put(87.33,15.33){\circle{2.11}}
\put(92.67,15.33){\circle{2.11}}
\put(97.67,15.33){\circle{2.11}}
\put(102.67,15.33){\circle{2.11}}
\put(73.00,10.67){\circle*{2.11}}
\put(77.33,10.67){\circle*{2.11}}
\put(92.67,10.67){\circle*{2.11}}
\put(97.67,10.67){\circle*{2.11}}
\put(102.67,10.67){\circle*{2.11}}
\put(107.67,10.67){\circle*{2.11}}
\put(29.67,28.00){\makebox(0,0)[cc]{$d_1=3$, q=1, s=1}}
\put(29.67,13.00){\makebox(0,0)[cc]{$d_2=6$, q=1, s=1}}
\put(65.00,33.00){\line(1,0){45.00}}
\put(110.00,33.00){\line(0,-1){25.00}}
\put(110.00,8.00){\line(-1,0){45.00}}
\put(65.00,28.00){\line(1,0){45.00}}
\put(65.00,13.00){\line(1,0){45.00}}
\put(75.00,33.00){\line(0,-1){10.00}}
\put(75.00,18.00){\line(0,-1){10.00}}
\put(85.00,32.67){\line(0,-1){9.67}}
\put(85.00,18.00){\line(0,-1){10.00}}
\put(65.00,22.67){\line(1,0){45.00}}
\put(65.00,18.00){\line(1,0){45.00}}
\end{picture}

$$ds^2=(H^{(1)}U^{(2)})^{1\over3}(H^{(2)}U^{(1)})^{2\over3}
[-(H^{(1)}U^{(1)}H^{(2)}U^{(2)})^{-1}dy^2+$$
$$+(H^{(1)}U^{(1)})^{-1}dy_1^2+ (H^{(1)}H^{(2)})^{-1}dy_2^2+$$
\begin{equation}
+(U^{(1)}U^{(2)})^{-1}dy_3^2+(H^{(2)}U^{(2)})^{-1}dy_4^2+
(H^{(2)}U^{(1)})^{-1}(dy_5^2+dy_6^2+dy_7^2)+ dx^{\alpha}dx^{\alpha}]
\end{equation}

%%%%%%%%%%%%%%%%%%%%%%%%%%%%%%%%%%%        END OF 6_3

\subsubsection {D=10}

Now we will investigate the case $D=10$ with two
fields.

 1)  {\it ''electric'' ansatz.}

  In this case we have the following condition
 $$ {\alpha\beta\over 2}+q-{d_1d_2\over
 8}+\sum_{i=1}^Nr_i\Delta_{ai}^{(1)}\Delta_{a'i}^{(2)}=0$$
 where all values are integer. Therefore there are following types of
 solutions.
  $$~$$

%\input{pic10.pic}
%%%%%%%%%%%%%%%%%%%%%%%%%%%%%%%%%%%          PIC10.PIC
\special{em:linewidth 0.4pt}
\unitlength 1.00mm
\linethickness{0.4pt}
\begin{picture}(110.00,25.00)(0.00,105.00)
\emline{70.00}{105.00}{1}{70.00}{130.00}{2}
\emline{75.00}{130.00}{3}{75.00}{120.00}{4}
\emline{75.00}{115.00}{5}{75.00}{105.00}{6}
\emline{85.00}{130.00}{7}{85.00}{120.00}{8}
\emline{85.00}{115.00}{9}{85.00}{105.00}{10}
\emline{95.00}{130.00}{11}{95.00}{120.00}{12}
\emline{95.00}{115.00}{13}{95.00}{105.00}{14}
\emline{105.00}{130.00}{15}{105.00}{120.00}{16}
\emline{105.00}{115.00}{17}{105.00}{105.00}{18}
\put(72.67,127.67){\circle{2.11}}
\put(77.67,127.67){\circle{2.11}}
\put(82.33,127.67){\circle{2.11}}
\put(72.67,122.67){\circle{2.11}}
\put(87.67,122.67){\circle{2.11}}
\put(92.33,122.67){\circle{2.11}}
\put(72.67,112.33){\circle{2.11}}
\put(97.67,112.33){\circle{2.11}}
\put(102.33,112.33){\circle{2.11}}
\put(77.67,112.33){\circle{2.11}}
\put(87.67,112.33){\circle{2.11}}
\put(72.67,107.67){\circle{2.11}}
\put(97.67,107.67){\circle{2.11}}
\put(102.33,107.67){\circle{2.11}}
\put(82.67,107.67){\circle{2.11}}
\put(92.33,107.67){\circle{2.11}}
\put(40.33,125.00){\makebox(0,0)[cc]{d=3, $\alpha=\pm{1\over 2}$, q=1, s=1}}
\put(40.00,110.00){\makebox(0,0)[cc]{d=5, $\beta=\mp{1\over 2}$, q=1, s=1}}
\emline{70.00}{130.00}{19}{110.00}{130.00}{20}
\emline{110.00}{130.00}{21}{110.00}{105.00}{22}
\emline{70.00}{125.00}{23}{110.00}{125.00}{24}
\emline{70.00}{110.00}{25}{110.00}{110.00}{26}
\emline{70.00}{105.00}{27}{110.00}{105.00}{28}
\emline{80.00}{130.00}{29}{80.00}{120.00}{30}
\emline{80.00}{115.00}{31}{80.00}{105.00}{32}
\emline{90.00}{130.00}{33}{90.00}{120.00}{34}
\emline{90.00}{115.00}{35}{90.00}{105.00}{36}
\emline{70.00}{120.00}{37}{110.00}{120.00}{38}
\emline{70.00}{115.00}{39}{110.00}{115.00}{40}
\end{picture}

$$ds^2=(H_1^{(1)}H_2^{(1)})^{3\over8}(H_1^{(2)}H_2^{(2)})^{5\over8}[-
(H_1^{(1)}H_2^{(1)}H_1^{(2)}H_2^{(2)})^{-1}dy^2+
(H_1^{(1)}H_1^{(2)})^{-1}dy_1^2+$$
$$+(H_1^{(1)}H_2^{(2)})^{-1}dy_2^2+
+(H_2^{(1)}H_1^{(2)})^{-1}dy_3^2+(H_2^{(1)}H_2^{(2)})^{-1}dy_4^2+$$
\begin{equation}
\label{D10f2e}
+(H_1^{(2)}H_2^{(2)})^{-1}(dy_5^2+dy_6^2)+dx^{\alpha}dx^{\alpha}]
\end{equation}

\begin{equation}
{\cal A}_8=4\pi L^6\sum_{c}(Q^{(1)}_{1c}Q^{(1)}_{2c}Q^{(2)}_{1c}
Q^{(2)}_{2c})^{1/2}
\end{equation}

%%%%%%%%%%%%%%%%%%%%%%%%%%%%%%%%%%%%%%%         END OF 10

There are many solutions with two fieldes when each of them has only one
electric or magnetic component. For example

%\input{pic14.pic}
%%%%%%%%%%%%%%%%%%%%%%%%%%%%%%%%%%            PIC14.PIC
\special{em:linewidth 0.4pt}
\unitlength 1.00mm
\linethickness{0.4pt}
\begin{picture}(100.00,15.00)(0.00,10.00)
\emline{60.00}{23.00}{1}{100.00}{23.00}{2}
\emline{100.00}{23.00}{3}{100.00}{8.00}{4}
\emline{100.00}{8.00}{5}{60.00}{8.00}{6}
\emline{60.00}{8.00}{7}{60.00}{23.00}{8}
\emline{80.00}{23.00}{9}{80.00}{18.00}{10}
\emline{80.00}{13.00}{11}{80.00}{8.00}{12}
\put(62.67,20.67){\circle{2.11}}
\put(67.33,20.67){\circle{2.11}}
\put(72.67,20.67){\circle{2.11}}
\put(62.67,10.33){\circle{2.11}}
\put(67.33,10.33){\circle{2.11}}
\put(77.67,10.33){\circle{2.11}}
\put(31.00,20.67){\makebox(0,0)[cc]{$d_1=3,\;\alpha=\pm{\sqrt{7}\over 2}$}}
\put(30.67,10.33){\makebox(0,0)[cc]{$d_2=3,\;\beta=\mp{\sqrt{7}\over 2}$}}
\emline{60.00}{18.00}{13}{100.00}{18.00}{14}
\emline{60.00}{13.00}{13}{100.00}{13.00}{14}
\emline{70.00}{23.00}{15}{70.00}{18.00}{16}
\emline{70.00}{13.00}{17}{70.00}{8.00}{18}
\emline{75.00}{23.00}{19}{75.00}{18.00}{20}
\emline{75.00}{13.00}{21}{75.00}{8.00}{22}
\emline{60.00}{18.00}{23}{100.00}{18.00}{24}
\end{picture}

\begin{equation}
\label{D10em4}
ds^2=(H_1H_2)^{3\over 11}[(H_1H_2)^{-{8\over 11}}(-dy^2+dy_1^2)
+H_1^{-{8\over 11}}dy_2^2+H_2^{-{8\over 11}}dy_3^2+dx^\alpha dx^\alpha]
\end{equation}
Centers of harmonic functions are singularities.
A family of such solutions was discussed in ~\cite{PapTow2}

%%%%%%%%%%%%%%%%%%%%%%%%%%%%%            END OF PIC14

  2) {\it ''magnetic'' ansatz.}
In this case we have the following condition
 $$ {\alpha\beta\over 2}+s-{d_1d_2\over
 8}+\sum_{i=1}^Nr_i\Lambda_{bi}^{(1)}\Lambda_{b'i}^{(2)}=0$$
 where all values are integer. Therefore there are three types of solutions.
 $$~$$
%\input{pic11.pic}
  %%%%%%%%%%%%%%%%%%%%%%%%%%%%%%%%%%%%       PIC11.PIC
\special{em:linewidth 0.4pt}
\unitlength 1.00mm
\linethickness{0.4pt}
\begin{picture}(110.00,20.00)(0.00,105.00)
\emline{70.00}{105.00}{1}{70.00}{130.00}{2}
\emline{75.00}{130.00}{3}{75.00}{120.00}{4}
\emline{75.00}{115.00}{5}{75.00}{105.00}{6}
\emline{85.00}{130.00}{7}{85.00}{120.00}{8}
\emline{85.00}{115.00}{9}{85.00}{105.00}{10}
\emline{95.00}{130.00}{11}{95.00}{120.00}{12}
\emline{95.00}{115.00}{13}{95.00}{105.00}{14}
\emline{105.00}{130.00}{15}{105.00}{120.00}{16}
\emline{105.00}{115.00}{17}{105.00}{105.00}{18}
\put(97.67,127.67){\circle*{2.11}}
\put(102.33,127.67){\circle*{2.11}}
\put(107.67,127.67){\circle*{2.11}}
\put(107.67,122.67){\circle*{2.11}}
\put(87.67,122.67){\circle*{2.11}}
\put(92.33,122.67){\circle*{2.11}}
\put(77.67,112.67){\circle*{2.11}}
\put(82.33,112.67){\circle*{2.11}}
\put(92.33,112.67){\circle*{2.11}}
\put(102.33,112.67){\circle*{2.11}}
\put(107.67,112.67){\circle*{2.11}}
\put(77.67,107.67){\circle*{2.11}}
\put(82.33,107.67){\circle*{2.11}}
\put(87.67,107.67){\circle*{2.11}}
\put(97.67,107.67){\circle*{2.11}}
\put(107.67,107.67){\circle*{2.11}}
\put(40.00,125.00){\makebox(0,0)[cc]{d=3, $\alpha=\pm{1\over 2}$, q=1, s=1}}
\put(39.67,110.00){\makebox(0,0)[cc]{d=5, $\beta=\mp{1\over 2}$, q=1, s=1}}
\emline{70.00}{130.00}{19}{110.00}{130.00}{20}
\emline{110.00}{130.00}{21}{110.00}{105.00}{22}
\emline{110.00}{105.00}{23}{70.33}{105.00}{24}
\emline{70.00}{125.00}{25}{110.00}{125.00}{26}
\emline{70.00}{110.00}{27}{110.00}{110.00}{28}
\emline{70.00}{120.00}{29}{110.00}{120.00}{30}
\emline{70.00}{115.00}{31}{110.00}{115.00}{32}
\end{picture}

$$ds^2=(U_1^{(1)}U_2^{(1)})^{5\over8}(U_1^{(2)}U_2^{(2)})^{3\over8}
[-(U_1^{(1)}U_2^{(1)}U_1^{(2)}U_2^{(2)})^{-1}dy^2+
(U_1^{(1)}U_2^{(1)})^{-1}(dy_1^2+$$
$$+dy_2^2)+(U_1^{(1)}U_1^{(2)})^{-1}dy_3^2+(U_1^{(1)}U_2^{(2)})^{-1}dy_4^2+$$
\begin{equation}
\label{D10f2m}
+(U_2^{(1)}U_1^{(2)})^{-1}dy_5^2+(U_2^{(1)}U_2^{(2)})^{-1}dy_6^2+
dx^{\alpha}dx^{\alpha}]
\end{equation}

\begin{equation}
{\cal A}_8=4\pi L^6\sum_{c}(P^{(1)}_{1c}P^{(1)}_{2c}P^{(2)}_{1c}
P^{(2)}_{2c})^{1/2}
\end{equation}

%%%%%%%%%%%%%%%%%%%%%%%%%%%%            END OF 11

 3){\it ''electric'' $+$ ''magnetic''}
ansatz.  In this case we have the additional conditions
$${\alpha\beta\over 2}-{d_1d_2\over
8}+\sum_{i=1}^Nr_i\Lambda_{bi}^{(1)}\Delta_{ai}^{(2)}=0,\;
{\alpha\beta\over 2}-{d_1d_2\over
8}+\sum_{i=1}^Nr_i\Lambda_{bi}^{(2)}\Delta_{ai}^{(1)}=0 $$

 We have three types of solutions.
  $$~$$
%\input{pic12_1.pic}

%%%%%%%%%%%%%%%%%%%%%%%%%%%%%%%%%%            PIC12_1.PIC
\special{em:linewidth 0.4pt}
\unitlength 1.00mm
\linethickness{0.4pt}
\begin{picture}(110.00,25.00)(0.00,5.00)
\emline{70.00}{8.00}{1}{70.00}{33.00}{2}
\emline{75.00}{33.00}{3}{75.00}{18.00}{4}
\emline{75.00}{13.00}{5}{75.00}{8.00}{6}
\emline{85.00}{33.00}{7}{85.00}{18.00}{8}
\emline{85.00}{13.00}{9}{85.00}{8.00}{10}
\emline{95.00}{33.00}{11}{95.00}{18.00}{12}
\emline{95.00}{13.00}{13}{95.00}{8.00}{14}
\emline{105.00}{33.00}{15}{105.00}{18.00}{16}
\emline{105.00}{13.00}{17}{105.00}{8.00}{18}
\put(72.67,30.67){\circle{2.11}}
\put(87.33,30.67){\circle{2.11}}
\put(97.33,30.67){\circle{2.11}}
\put(72.67,25.67){\circle{2.11}}
\put(92.33,25.67){\circle{2.11}}
\put(103.00,25.67){\circle{2.11}}
\put(107.67,20.67){\circle*{2.11}}
\put(103.00,20.67){\circle*{2.11}}
\put(97.33,20.67){\circle*{2.11}}
\put(72.67,10.33){\circle{2.11}}
\put(77.67,10.33){\circle{2.11}}
\put(82.33,10.33){\circle{2.11}}
\put(97.33,10.33){\circle{2.11}}
\put(103.00,10.33){\circle{2.11}}
\put(40.00,25.67){\makebox(0,0)[cc]{d=3, $\alpha=\pm{1\over 2}$, q=1, s=1}}
\put(39.67,10.33){\makebox(0,0)[cc]{d=5, $\beta=\mp{1\over 2}$, q=1, s=1}}
\emline{70.00}{33.00}{19}{110.00}{33.00}{20}
\emline{110.00}{33.00}{21}{110.00}{8.00}{22}
\emline{110.00}{8.00}{23}{70.00}{8.00}{24}
\emline{70.00}{28.00}{25}{110.00}{28.00}{26}
\emline{70.00}{23.00}{27}{110.00}{23.00}{28}
\emline{90.00}{33.00}{29}{90.00}{18.00}{30}
\emline{90.00}{13.00}{31}{90.00}{8.00}{32}
\emline{100.00}{33.00}{33}{100.00}{18.00}{34}
\emline{100.00}{13.00}{35}{100.00}{8.00}{36}
\emline{70.00}{18.00}{37}{110.00}{18.00}{38}
\emline{70.00}{13.00}{39}{110.00}{13.00}{40}
\end{picture}

$$ds^2=(H_1^{(1)}H_2^{(1)})^{3\over8}(UH^{(2)})^{5\over8}
[-(H_1^{(1)}H_2^{(1)}UH^{(2)})^{-1}dy^2+(H^{(2)}U)^{-1}(dy_1^2+dy_2^2)+$$
$$+(H_1^{(1)}U)^{-1}dy_3^2+(H_2^{(1)}U)^{-1}dy_4^2+$$
\begin{equation}
\label{D10f2em1}
+(H_1^{(1)}H^{(2)})^{-1}dy_5^2+(H_2^{(1)}H^{(2)})^{-1}dy_6^2
+dx^{\alpha}dx^{\alpha}]
\end{equation}

%\begin{equation}
%{\cal A}_2=4\pi\sum_{c_0}(Q^{(1)}_{1c_0}Q^{(1)}_{2c_0}Q^{(2)}_{c_0}
%P_{c_0})^{1/2}
%\end{equation}

%%%%%%%%%%%%%%%%%%%%%%%       END OF 12_1

%\input{pic12_2.pic}

%%%%%%%%%%%%%%%%%%%%%%%%%%%%%%%%%%%           PIC12_2.PIC
\special{em:linewidth 0.4pt}
\unitlength 1.00mm
\linethickness{0.4pt}
\begin{picture}(110.00,25.00)(0.00,10.00)
\emline{70.00}{8.00}{1}{70.00}{33.00}{2}
\emline{75.00}{33.00}{3}{75.00}{18.00}{4}
\emline{75.00}{13.00}{5}{75.00}{8.00}{6}
\emline{85.00}{33.00}{7}{85.00}{18.00}{8}
\emline{85.00}{13.00}{9}{85.00}{8.00}{10}
\emline{95.00}{33.00}{11}{95.00}{18.00}{12}
\emline{95.00}{13.00}{13}{95.00}{8.00}{14}
\emline{105.00}{33.00}{15}{105.00}{18.00}{16}
\emline{105.00}{13.00}{17}{105.00}{8.00}{18}
\put(72.67,30.67){\circle{2.11}}
\put(77.67,30.67){\circle{2.11}}
\put(82.33,30.67){\circle{2.11}}
\put(87.33,30.67){\circle{2.11}}
\put(92.33,30.67){\circle{2.11}}
\put(72.67,25.67){\circle{2.11}}
\put(77.67,25.67){\circle{2.11}}
\put(82.33,25.67){\circle{2.11}}
\put(97.33,25.67){\circle{2.11}}
\put(103.00,25.67){\circle{2.11}}
\put(107.67,20.67){\circle*{2.11}}
\put(97.33,20.67){\circle*{2.11}}
\put(87.33,20.67){\circle*{2.11}}
\put(77.67,20.67){\circle*{2.11}}
\put(82.33,20.67){\circle*{2.11}}
\put(72.67,10.33){\circle{2.11}}
\put(87.33,10.33){\circle{2.11}}
\put(97.33,10.33){\circle{2.11}}
\put(40.00,28.00){\makebox(0,0)[cc]{d=5, $\alpha=\pm{1\over 2}$, q=1, s=1}}
\put(39.67,10.33){\makebox(0,0)[cc]{d=3, $\beta=\mp{1\over 2}$, q=1, s=1}}
\emline{70.00}{33.00}{19}{110.00}{33.00}{20}
\emline{110.00}{33.00}{21}{110.00}{7.67}{22}
\emline{110.00}{8.00}{23}{70.00}{8.00}{24}
\emline{70.00}{28.00}{25}{110.00}{28.00}{26}
\emline{70.00}{23.00}{27}{110.00}{23.00}{28}
\emline{100.00}{33.00}{29}{100.00}{18.00}{30}
\emline{100.00}{13.00}{31}{100.00}{8.00}{32}
\emline{70.00}{17.66}{33}{110.00}{17.66}{34}
\emline{70.00}{13.00}{35}{110.00}{13.00}{36}
\end{picture}

$$ds^2=(H_1^{(1)}H_2^{(1)})^{5\over8}(UH^{(2)})^{3\over8}
[-(H_1^{(1)}H_2^{(1)}UH^{(2)})^{-1}dy^2+(H_1^{(1)}H_2^{(1)})^{-1}(dy_1^2+$$
$$+dy_2^2)+(H_1^{(1)}H^{(2)})^{-1}dy_3^2+(H_1^{(1)}U)^{-1}dy_4^2+$$
\begin{equation}
\label{D10f2em2}
+(H_2^{(1)}H^{(2)})^{-1}dy_5^2+(H_2^{(1)}U)^{-1}dy_6^2+
dx^{\alpha}dx^{\alpha}]
\end{equation}

%\begin{equation}
%{\cal A}_2=4\pi\sum_{c_0}(Q^{(1)}_{1c_0}Q^{(1)}_{2c_0}Q^{(2)}_{c_0}
%P_{c_0})^{1/2}
%\end{equation}

%%%%%%%%%%%%%%%%%%%%%%%%%%%%%          END OF 12_2

%\input{pic12_3.pic}

%%%%%%%%%%%%%%%%%%%%%%%%%%%%%%%%           PIC12_3.PIC
\special{em:linewidth 0.4pt}
\unitlength 1.00mm
\linethickness{0.4pt}
\begin{picture}(110.00,25.33)(0.00,10.00)
\emline{70.00}{8.33}{1}{70.00}{33.33}{2}
\emline{75.00}{33.33}{3}{75.00}{23.33}{4}
\emline{75.00}{18.33}{5}{75.00}{8.33}{6}
\emline{85.00}{33.33}{7}{85.00}{23.33}{8}
\emline{85.00}{18.33}{9}{85.00}{8.33}{10}
\emline{95.00}{33.33}{11}{95.00}{23.33}{12}
\emline{95.00}{18.33}{13}{95.00}{8.33}{14}
\emline{105.00}{33.33}{15}{105.00}{23.33}{16}
\emline{105.00}{18.33}{17}{105.00}{8.33}{18}
\put(72.67,31.00){\circle{2.11}}
\put(87.33,31.00){\circle{2.11}}
\put(97.33,31.00){\circle{2.11}}
\put(72.67,15.66){\circle{2.11}}
\put(77.67,15.66){\circle{2.11}}
\put(82.33,15.66){\circle{2.11}}
\put(87.33,15.66){\circle{2.11}}
\put(92.33,15.66){\circle{2.11}}
\put(87.33,26.00){\circle*{2.11}}
\put(92.33,26.00){\circle*{2.11}}
\put(107.67,26.00){\circle*{2.11}}
\put(107.67,11.00){\circle*{2.11}}
\put(77.67,11.00){\circle*{2.11}}
\put(82.33,11.00){\circle*{2.11}}
\put(87.33,11.00){\circle*{2.11}}
\put(97.33,11.00){\circle*{2.11}}
\put(39.67,28.33){\makebox(0,0)[cc]{d=3, $\alpha=\pm{1\over 2}$, q=1, s=1}}
\put(39.67,13.33){\makebox(0,0)[cc]{d=5, $\beta=\mp{1\over 2}$, q=1, s=1}}
\emline{70.00}{33.33}{19}{110.00}{33.33}{20}
\emline{110.00}{33.33}{21}{110.00}{8.33}{22}
\emline{110.00}{8.33}{23}{70.00}{8.33}{24}
\emline{70.33}{28.33}{25}{110.00}{28.33}{26}
\emline{70.00}{13.33}{27}{110.00}{13.33}{28}
\emline{90.00}{33.33}{29}{90.00}{23.33}{30}
\emline{90.00}{18.33}{31}{90.00}{8.33}{32}
\emline{100.00}{33.33}{33}{100.00}{23.33}{34}
\emline{100.00}{18.33}{35}{100.00}{8.33}{36}
\emline{70.00}{23.33}{37}{110.00}{23.33}{38}
\emline{70.00}{18.33}{39}{110.00}{18.33}{40}
\end{picture}

$$ds^2=(H^{(1)}U^{(2)})^{3\over8}(H^{(2)}U^{(1)})^{5\over8}
[-(H^{(1)}U^{(1)}H^{(2)}U^{(2)})^{-1}dy^2+
(H^{(2)}U^{(1)})^{-1}(dy_1^2+dy_2^2)+$$
$$+(H^{(1)}H^{(2)})^{-1}dy_3^2+(H^{(2)}U^{(2)})^{-1}dy_4^2+$$
\begin{equation}
\label{D10f2em3}
+(H^{(1)}U^{(1)})^{-1}dy_5^2+(U^{(1)}U^{(2)})^{-1}dy_6^2+
dx^{\alpha}dx^{\alpha}]
\end{equation}

%\begin{equation}
%{\cal A}_2=4\pi\sum_{c_0}(Q^{(1)}_{c_0}P^{(1)}_{c_0}Q^{(2)}_{c_0}
%P^{(2)}_{c_0})^{1/2}
%\end{equation}
%%%%%%%%%%%%%%%%%%%%%%%%%%%%%%%%%         END OF 12_3

\newpage

\section{Depending Functions}

Here we consider the ansatzes (\ref{1.1}) and (\ref{1.2})
with
functions $C_{a}$ and $\chi_{b}$ which are not independent.
We are not going to present a general consideration since it is
straightforward but rather involved. We construct only special
solutions with depending functions $C_{a}$ and $\chi_{b}$.  Let us consider
the theory (\ref{1}) with $\alpha=0$,
'electric' $+$ 'magnetic' ansatz with equal number of branches
$(E=M)$ and $C_a=\chi_a$. Substituting
expressions for $A,\;B,$ and $F_i$ in 'no-force' condition for
diagonal part one can obtain \begin{equation} \label{nf1}
-h_a^2td-v_a^2ud+v_a^2\frac{1}{2}\sum_{i=1}^Nr_i\Delta_{ai}\Lambda_{ai}+1=0
\end{equation}
\begin{equation}
\label{nf2}
-h_a^2ud-v_a^2td+h_a^2\frac{1}{2}\sum_{i=1}^Nr_i\Delta_{ai}\Lambda_{ai}+1=0
\end{equation}

These equations are consistent under an assumption
 \begin{equation}
 \label{hv}
 v_a^2=h_a^2
\end{equation}
Then we have
  \begin{equation}
   \label{h}
  h_a^2={2\over d-\sum_{i=1}^Nr_i\Delta_{ai}\Lambda_{ai}}
\end{equation}
 A non-diagonal part of the 'no-force' conditions gives
  \begin{equation}
  \label{nd1}
\sum_{i=1}^Nr_i\Delta_{a'i}\Lambda_{ai}-\sum_{i=1}^Nr_i\Delta_{a'i}
\Delta_{ai}-q=0
 \end{equation}
\begin{equation}
   \label{nd2}
\sum_{i=1}^Nr_i\Delta_{ai}\Lambda_{a'i}-\sum_{i=1}^Nr_i\Lambda_{a'i}
\Lambda_{ai}-s=0
 \end{equation}
where $a'\not=a$. Note that there is no explicit dependence on D in the
conditions (\ref{nd1}), (\ref{nd2}).

Some examples.
For $D=11,\;d=4,\;q=s=1$ we can put
$$\sum_{i=1}^Nr_i\Delta_{ai}\Lambda_{a'i}=1,\;
\sum_{i=1}^Nr_i\Delta_{a'i} \Delta_{ai}=0$$
$$\sum_{i=1}^Nr_i\Lambda_{a'i}
\Lambda_{ai}=0,\;\sum_{i=1}^Nr_i\Delta_{ai}\Lambda_{ai}=0$$

%\input{pic16.pic}

%%%%%%%%%%%%%%%%%%%%%%%%%%%%%%%%          PIC16.PIC
\unitlength 1.00mm
\linethickness{0.4pt}
\begin{picture}(110.00,40.00)(0.00,5.00)
\put(65.00,25.00){\line(0,1){15.00}}
\put(65.00,40.00){\line(1,0){5.33}}
\put(70.33,40.00){\line(-1,0){5.33}}
\put(65.00,40.00){\line(1,0){5.33}}
\put(70.33,40.00){\line(-1,0){5.33}}
\put(70.00,40.00){\line(0,-1){15.00}}
\put(67.67,37.33){\circle{2.11}}
\put(67.67,32.67){\circle{2.11}}
\put(67.67,27.33){\circle{2.11}}
\put(80.00,40.00){\line(0,-1){15.00}}
\put(90.00,40.00){\line(0,-1){15.00}}
\put(100.00,40.00){\line(0,-1){15.00}}
\put(77.33,37.33){\circle{2.11}}
\put(82.67,32.67){\circle{2.11}}
\put(87.00,32.67){\circle{2.11}}
\put(93.00,27.33){\circle{2.11}}
\put(97.00,27.33){\circle{2.11}}
\put(72.67,37.33){\circle{2.11}}
\put(29.67,25.00){\makebox(0,0)[cc]{d=3, q=1, s=1}}
\put(70.00,40.00){\line(1,0){40.00}}
\put(110.00,40.00){\line(0,-1){15.00}}
\put(110.00,25.00){\line(-1,0){45.00}}
\put(65.00,35.00){\line(1,0){45.00}}
\put(65.00,30.00){\line(1,0){45.00}}
\put(65.00,25.00){\line(0,-1){15.00}}
\put(65.00,10.00){\line(1,0){45.00}}
\put(110.00,10.00){\line(0,1){15.00}}
\put(65.00,20.00){\line(1,0){45.00}}
\put(65.00,15.00){\line(1,0){45.00}}
\put(70.00,25.00){\line(0,-1){15.00}}
\put(90.00,25.00){\line(0,-1){15.00}}
\put(100.00,25.00){\line(0,-1){15.00}}
\put(105.00,40.00){\line(0,-1){30.00}}
\put(80.00,25.00){\line(0,-1){15.00}}
\put(107.33,22.67){\circle*{2.11}}
\put(82.67,22.67){\circle*{2.11}}
\put(93.00,22.67){\circle*{2.11}}
\put(107.33,17.67){\circle*{2.11}}
\put(72.67,17.67){\circle*{2.11}}
\put(97.00,17.67){\circle*{2.11}}
\put(107.33,12.67){\circle*{2.11}}
\put(77.33,12.67){\circle*{2.11}}
\put(87.00,12.67){\circle*{2.11}}
\end{picture}

%%%%%%%%%%%%%%%%%%%%%%%%%%%%%%%%%%           END OF 16

$$ds^2=
(H_1H_2H_3)^{2\over3}[-(H_1H_2H_3)^{-{4\over3}}dy^2+(H_1^2H_3)^{-{2/3}}dy_1^2
+(H_1^{2}H_2)^{-{2/3}}dy_2^2+(H_2^2H_3)^{-{2/3}}dy_3^2+$$
 \begin{equation}
 \label{cm}
+(H_1H_2^2)^{-{2/3}}dy_4^2+(H_2H_3^2)^{-{2/3}}dy_5^2+
(
H_1H_3^2)^{-{2/3}}dy_6^2+(H_1H_2H_3)^{-{2/3}}dy^{2}_{7}
+dx^\alpha dx^\alpha]
  \end{equation}
For the area of horizon we have
 \begin{equation}
 {\cal A}_9=4\pi L^7(H_1H_2H_3)^{2/3}
 \end{equation}

Another example. Let us consider
 $D=10, d=4, q=s=1$

$$\sum_{i=1}^Nr_i\Delta_{ai}\Lambda_{a'i}=2,\;
\sum_{i=1}^Nr_i\Delta_{a'i} \Delta_{ai}=1$$
$$\sum_{i=1}^Nr_i\Lambda_{a'i}
\Lambda_{ai}=1,\;\sum_{i=1}^Nr_i\Delta_{ai}\Lambda_{ai}=0$$

%\input{pic15.pic}

%%%%%%%%%%%%%%%%%%%%%%%%%%%%%%%%%         PIC15.PIC
\special{em:linewidth 0.4pt}
\unitlength 1mm
\linethickness{0.4pt}
\begin{picture}(110.33,48.00)
\emline{70.00}{8.00}{1}{110.00}{8.00}{2}
\emline{110.00}{8.00}{3}{110.00}{48.00}{4}
\emline{110.00}{48.00}{5}{70.00}{48.00}{6}
\emline{70.00}{48.00}{7}{70.00}{8.00}{8}
\emline{70.00}{43.00}{9}{110.00}{43.00}{10}
\emline{70.00}{38.00}{11}{110.00}{38.00}{12}
\emline{70.00}{33.00}{13}{110.00}{33.00}{14}
\emline{70.00}{28.00}{15}{110.00}{28.00}{16}
\emline{70.00}{23.00}{17}{110.33}{23.00}{18}
\emline{70.00}{18.00}{19}{110.00}{18.00}{20}
\emline{70.00}{13.00}{21}{110.00}{13.00}{22}
\emline{75.00}{48.00}{23}{75.00}{7.67}{24}
\emline{80.00}{48.00}{25}{80.00}{8.00}{26}
\emline{85.00}{48.00}{27}{85.00}{8.00}{28}
\emline{90.00}{48.00}{29}{90.00}{8.00}{30}
\emline{95.00}{48.00}{31}{95.00}{8.00}{32}
\emline{100.00}{48.00}{33}{100.00}{8.00}{34}
\emline{105.00}{48.00}{35}{105.00}{8.00}{36}
\put(72.67,45.33){\circle{2.11}}
\put(77.67,45.33){\circle{2.11}}
\put(82.67,45.33){\circle{2.11}}
\put(87.67,45.33){\circle{2.11}}
\put(72.67,40.67){\circle{2.11}}
\put(77.67,40.67){\circle{2.11}}
\put(92.33,40.67){\circle{2.11}}
\put(97.67,40.67){\circle{2.11}}
\put(72.67,35.67){\circle{2.11}}
\put(82.67,35.67){\circle{2.11}}
\put(92.33,35.67){\circle{2.11}}
\put(102.67,35.67){\circle{2.11}}
\put(72.67,30.67){\circle{2.11}}
\put(87.67,30.67){\circle{2.11}}
\put(97.67,30.67){\circle{2.11}}
\put(102.67,30.67){\circle{2.11}}
\put(107.33,25.67){\circle*{2.11}}
\put(92.33,25.33){\circle*{2.11}}
\put(97.67,25.33){\circle*{2.11}}
\put(102.67,25.33){\circle*{2.11}}
\put(107.33,20.67){\circle*{2.11}}
\put(102.67,20.67){\circle*{2.11}}
\put(87.67,20.67){\circle*{2.11}}
\put(82.67,20.67){\circle*{2.11}}
\put(107.33,15.67){\circle*{2.11}}
\put(97.67,15.67){\circle*{2.11}}
\put(87.67,15.67){\circle*{2.11}}
\put(77.67,15.67){\circle*{2.11}}
\put(77.67,10.67){\circle*{2.11}}
\put(82.67,10.67){\circle*{2.11}}
\put(92.33,10.67){\circle*{2.11}}
\put(107.33,10.67){\circle*{2.11}}
\put(40.00,28.00){\makebox(0,0)[cc]{$d=4,\;\alpha=0,\;q=1,\;s=1$}}
\end{picture}

%%%%%%%%%%%%%%%%%%%%%%%%%%%%%%            END OF 15

We have the following metric
$$ds^2=(H_1H_2H_3H_4)^{1\over2}[-(H_1H_2H_3H_4)^{-1}dy^2+(H_1H_2)^{-1}dy_1^2
+(H_1H_3)^{-1}dy_2^2+(H_1H_4)^{-1}dy_3^2+$$
 \begin{equation}
 \label{cma}
    +(H_2H_3)^{-1}dy_4^2+(H_2H_4)^{-1}dy_5^2+(H_3H_4)^{-1}dy_6^2+dx^\alpha
dx^\alpha]
  \end{equation}
 This solution was discussed in \cite{klebanov}.

 \newpage
\section*{Acknowledgments}
We are grateful to I.V.Volovich for useful discussions and to
M.G.Ivanov for attracting our attention to special cases when the
stress-energy tensor is non-diagonal. This work is supported by the RFFI
grant 96-01-00608.


\begin{thebibliography}{99}
\bibitem{HT} C.Hull and P.Townsend, Nucl.Phys.B438(1995)109
\bibitem{W} E.Witten, Nucl.Phys.B443(1995)85
\bibitem{PT} P.Townsend, hep-th 9507048
\bibitem{JHS} J.H.Schwarz,hep-th/9510086, hep-th/9607201, hep-th/9607201

\bibitem{DGHR} A.Dubholkar, G.W.Gibbons, J.A.Harvey, F.Ruiz,
 Nucl.Phys. {\bf 340} (1990) 33.
\bibitem{Hor}  G.T.Horowitz and A.Strominger, Nucl.Phys. B360 (1991) 197.
\bibitem{duff1} M.J.Duff and K.S.Stelle, Phys.Lett.{\bf 253}(1991) 113.
\bibitem{guven}R.Guven, Phys.Lett.{\bf 276}(1992) 49,
 Phys.Lett.{\bf 212}(1988) 277.
\bibitem{Re} R.Kallosh, A.Linde, T.Ortin, A.Peet and A.van Proeyen,
 Phys.Rev. {\bf D46} (1992) 5278.
\bibitem{Lu} H.Lu, C.N.Pope, E.Sergin and K.Stelle,
 hep-th/9508042.
\bibitem{duff2} M.J.Duff, R.R.Khuri and J.X.Lu,
 Phys.Rep. {\bf 259} (1995) 213.
\bibitem{gibbons} G.W.Gibbons, G.T.Horowitz and P.K.Townsend,
 Class. Quant. Grav.{\bf 12}  (1995) 297.
\bibitem{Cvetic} M.Cvetic and D.Youm, hep-th/9510098
\bibitem{lu}H.Lu and C.N.Pope, hep-th/9512153; hep-th/9512012
\bibitem{Vafa} A.Strominger and C.Vafa, hep-th/9601029
\bibitem{Hor} G.T.Horowitz, gr-qc/ 9604051
\bibitem{CM} C.Callan and J.Maldacena, hep-th/9602043
\bibitem{TS}  A.A.Tseytlin, hep-th/9601177
\bibitem{ToPa1}  G.Papadopoulos and P.K.Townsend,
 hep-th/9603087
\bibitem{Ts1} A.A.Tseytlin, hep-th/9604035
\bibitem{kelly} H.Lu, C.Pope and K.Stelle, hep-th/9604058
\bibitem{Pap} G.Papadopoulos, hep-th/9604068.
\bibitem{Lip} G.Lifschytz, hep-th/9604156
\bibitem{klebanov} I.R.Klebanov and A.A.Tseytlin, hep-th/9604166
\bibitem{BB} K.Behrndt, E.Bergshoeff and B.Janssen, hep-th/9604168
\bibitem{Gaun}  J.P.Gauntlett, D.A.Kastor and J.Traschen, hep-th/9604179.
\bibitem{berg} E.Bergshoeff, R.Kallosh, T.Ortin,
                                   hep-th/9605059
\bibitem{Ts2} A. A. Tseytlin, gr-qc/9608044
\bibitem{AV} A.Volovich, hep-th/9608095
\bibitem{Berg} E.Bergshoeff, M.De Roo and S.Panda, hep-th/9609056
 \bibitem{PapTow2} G.Papadopoulos and P.K.Townsend, hep-th/9609095
\bibitem{Ts212}  A. A. Tseytlin, hep-th/9609212
\bibitem{AVV} I.Ya. Aref'eva, K.S.Viswanathan  and I.V.Volovich,
                        hep-th/9609225
\bibitem{Lust} K.Behrndt, G;L.Cardoso, B.de Wit, R.Kallosh, D.Lust
and T.Mohaupt, hep-th/9610105.
 \bibitem{Em} R. Emparan, hep-th/9610170.
\bibitem{kl} C.G.Callan, Jr., S.S.Gubser,I.R.Klebanov and
            A.A.Tseytlin, hep-th/9610172
\bibitem{IN}I.Ya. Aref'eva and  A.Volovich, hep-th/9611026
\end{thebibliography}
\end{document}